\documentclass[12pt]{article}

\usepackage[utf8]{inputenc}
\usepackage{times}
\usepackage{graphicx,amssymb}
\usepackage{subcaption}
\usepackage{siunitx}
\usepackage{amsmath}
\usepackage[normalem]{ulem} % for \sout
\usepackage{xcolor}
\DeclareFontFamily{U}{mathx}{\hyphenchar\font45}
\DeclareFontShape{U}{mathx}{m}{n}{
      <5> <6> <7> <8> <9> <10>
      <10.95> <12> <14.4> <17.28> <20.74> <24.88>
      mathx10
      }{}
\DeclareSymbolFont{mathx}{U}{mathx}{m}{n}
\DeclareFontSubstitution{U}{mathx}{m}{n}
\DeclareMathAccent{\widecheck}{0}{mathx}{"71}

\newcommand{\Yhat}{\widehat{Y}}

\newcommand{\Frac}[2]{{{#1}/{#2}}}  % an "inert" form of \frac
\newcommand{\beq}{\begin{equation}}
\newcommand{\eeq}{\end{equation}}
\newcommand{\beqan}{\begin{eqnarray*}}
\newcommand{\eeqan}{\end{eqnarray*}}

\newcommand{\openCase}  {\left\{ \begin{array}{@{\,}ll}}
\newcommand{\openCasell}{\left\{ \begin{array}{@{\,}ll}}
\newcommand{\openCasecl}{\left\{ \begin{array}{@{\,}cl}}
\newcommand{\openCaserl}{\left\{ \begin{array}{@{\,}rl}}
\newcommand{\openCaseTablell}{\left\{ \begin{array}{@{}ll}}
\newcommand{\openCaseTablecl}{\left\{ \begin{array}{@{}cl}}
\newcommand{\openCaseTablerl}{\left\{ \begin{array}{@{}rl}}
\newcommand{\closeCase} {\end{array} \right.}

\def\Mtilde{\widetilde{M}}

\def\Utilde{\widetilde{U}}

\def\Xtilde{\widetilde{X}}

\def\smid{\,|\,}  % Like \mid, but with less space; use for conditioning
\def\sMid{\,;\,}  % Version of \smid for non-Bayesian case
\newcommand{\iE}[1]{\mathrm{E}[\,{#1}\,]}   % Expectation classical (inline)
\newcommand{\var}[1]{\mathrm{var}\!\left({#1}\right)}   % Variance
\newcommand{\normal}[2]{{\mathcal{N}(#1,#2)}}  % normal distribution
\newcommand{\Poisson}[1]{{\mathrm{Poisson}(#1)}}  % Poisson distribution
\def\Utilde{{\widetilde{U}}}

\def\etaHat{\widehat{\eta}}

\def\etaICA{\widehat{\eta}_{\rm ICA}}
\def\etaTR{\widehat{\eta}_{\rm ICA}}
\def\etaconv{\widehat{\eta}_{\rm conv}}

\def\etaconv{\widehat{\eta}_{\rm conv}}
\def\Mtildecorr{\widetilde{M}_{\rm corr}}
\def\um{\si{\micro\meter}}
\def\mm{\si{\milli\meter}}

\def\us{\si{\micro\second}}
\def\ns{\si{\nano\second}}

\def\pA{\si{\pico\ampere}}

\def\keV{\si{\kilo\electronvolt}}

\def\V{\si{\volt}}

\def\cmu{{c_{\mu}}}
\def\csig{{c_{\sigma}}}

\usepackage{cleveref} % must come last

\begin{document}

\title{Shot noise-mitigated secondary electron imaging with ion count-aided microscopy}

\author{
        Akshay Agarwal$^{1\ast}$,
        Leila Kasaei$^{2}$,
        Xinglin He$^{1}$,\\
        Ruangrawee Kitichotkul$^{1}$,
        O\u{g}uz Ka\u{g}an Hitit$^{1}$,
        Minxu Peng$^{1}$,\\
        J. Albert Schultz$^{3}$,
        Leonard C. Feldman$^{2}$, and
        Vivek K Goyal$^{1\ast}$ \\
\normalsize{$^{1}$Department of Electrical and Computer Engineering, Boston University} \\
\normalsize{$^{2}$Department of Physics, Rutgers University} \\
\normalsize{$^{3}$Ionwerks Inc.} \\
\normalsize{$^\ast$To whom correspondence should be addressed: akshayag@bu.edu, v.goyal@ieee.org}
    }
\date{}

\maketitle

\abstract{
\textbf{Modern science is dependent on imaging on the nanoscale, often achieved through processes that detect secondary electrons created by a highly focused incident charged particle beam.
Multiple types of measurement noise limit the ultimate trade-off between the image quality and the incident particle dose, which can preclude useful imaging of dose-sensitive samples.
Existing methods to improve image quality do not fundamentally mitigate the noise sources.
Furthermore, barriers to assigning a physically meaningful scale make the images qualitative.
Here we introduce ion count-aided microscopy (ICAM),
which is a quantitative imaging technique that uses statistically principled estimation of the secondary electron yield.
With a readily implemented change in data collection, ICAM substantially reduces source shot noise.
In helium ion microscopy, we demonstrate 3$\times$ dose reduction and a good match between these empirical results and theoretical performance predictions.
ICAM facilitates imaging of fragile samples and may make imaging with heavier particles more attractive.}
}
%\VKGcomment{I feel that the ``thus'' is natural with the first part of what follows but not with the second part (imaging with heaver particles).}
%250 words max

\section*{Introduction}
Secondary electron imaging (SEI) modalities such as scanning electron microscopy (SEM)
and helium ion microscopy (HIM) are widely used in nanoscale characterization and analysis~\cite{Reimer1998,WardNE:06}. 
SEM is employed in diverse applications such as critical-dimension metrology and inspection for semiconductor devices~\cite{Vladar2001}, materials characterization in geology~\cite{Reed2005}, and examination of biological samples~\cite{Thiberge2004}. 
With its applicability to non-conducting materials (not requiring sample coating before imaging), HIM is especially useful in the high-resolution imaging of biological samples such as animal organs~\cite{Boden2012,Rice2013}, tumor cells~\cite{Bazou2011}, and viruses~\cite{Lepannen2017,Merolli2022}.
In SEI, a beam of charged particles (electrons or ions) is raster scanned across the sample being imaged.
At each scan location, the incident beam is held in place while it excites secondary electrons (SEs) that are detected by a secondary electron detector (SED). 
A detected signal intensity is converted to a
pixel brightness value.
Scanning over a rectangular grid forms the final image of the sample. 
%[The ideal]
Ideally, an SE image would be a map of the sample's SE yield $\eta$, \textit{i.e.}, the mean number of SEs generated per incident particle.
However, since SEDs do not have sufficient energy resolution to count SEs, and the gain and efficiency of the detector are generally unknown to the user, conventional SE images are merely qualitative---they cannot be directly mapped to pixelwise SE yields.

The quality of SEI is affected by three sources of noise:
random variation
in the number of incident particles (\emph{source shot noise}),
in the number of emitted SEs for each incident particle (\emph{target shot noise}),
and in the signal produced by the SED in response to SEs (\emph{detector noise}). 
These noise sources limit the image quality achievable in SEI at any given imaging dose.
For rugged materials such as gold or copper, the image quality can be improved by increasing dose. 
However, for fragile, radiation-sensitive materials, increase in dose also increases the damage imparted during imaging, such as atomic displacement and radiolysis~\cite{Joy1996b,Egerton2004}. Displacement damage (such as sputtering and defect creation) is especially significant for HIM as compared to SEM, and it is worse yet with gallium or neon beams~\cite{Wirtz2012}.

One approach to reducing the impact of detector noise is to count pulses in the SED voltage signal to infer the number of SEs. Such counting has been accomplished by external pulse counting circuits~\cite{Uchikawa1992} and in software~\cite{Joy2008,Agarwal:2023-U}.
While this technique improves imaging signal-to-noise-ratio at low SE yields, it
%[cannot be directly applied]
is ineffective at the higher yields achievable with ions~\cite{WardNE:06}, where multiple SEs may be detected simultaneously.
Post-processing methods to reduce noise,
such as Gaussian/Poisson deconvolution~\cite{Roels2016,Vanderlinde2007}, compressed sensing~\cite{Anderson2013,Kovarik2016,Stevens2018}, inpainting~\cite{Pang2021}, and adaptive scanning~\cite{Dahmen2016},
rely on assumptions about the underlying image structure rather than on
the statistics of the particle source or SE emission. 
Where applicable, these types of post-processing can be combined with more fundamental physics-based improvements.

In this work, we introduce ion count-aided microscopy
(ICAM), which uses novel data processing to nearly eliminate source shot noise. 
The target shot noise is inherent to the measured beam--sample interactions, and the detector noise can be reduced by altering the choice of detector technology;
thus, ICAM approaches the accuracy limits of SEI\@.
We demonstrate ICAM with a helium ion microscope to achieve quantitative, nanoscale SE yield metrology.
%one of the noise sources---the randomness in the number of incident ions. 
%\AAcomment{Line overflow}
Building upon a time-resolved measurement concept developed under an assumption of perfect SE observation~\cite{PengMBBG:20,PengMBG:21,Agarwal2023b},
we use the full waveform of the SED signal
to infer ion incidence events.
Using the number of observed ion incidence events improves SE yield estimation significantly;
we demonstrate reduction of the dose required for a given image quality by up to a factor of 3\@.
Theoretical analysis suggests that the dose reduction factor provided by ICAM is approximately equal to the SE yield $\eta$ and thus can be much larger for other incident particles and samples~\cite{Fehn1976}.
Although specialized for SEI, our work demonstrates improvements in imaging that can be accomplished with careful consideration of the statistical properties of the underlying data. 
It may inspire similar considerations in other types of imaging and materials characterization.

\section*{Results}

\subsection*{Data collection to allow incident ion counting}
\begin{figure*}
    \centering
    \includegraphics[width=\linewidth]{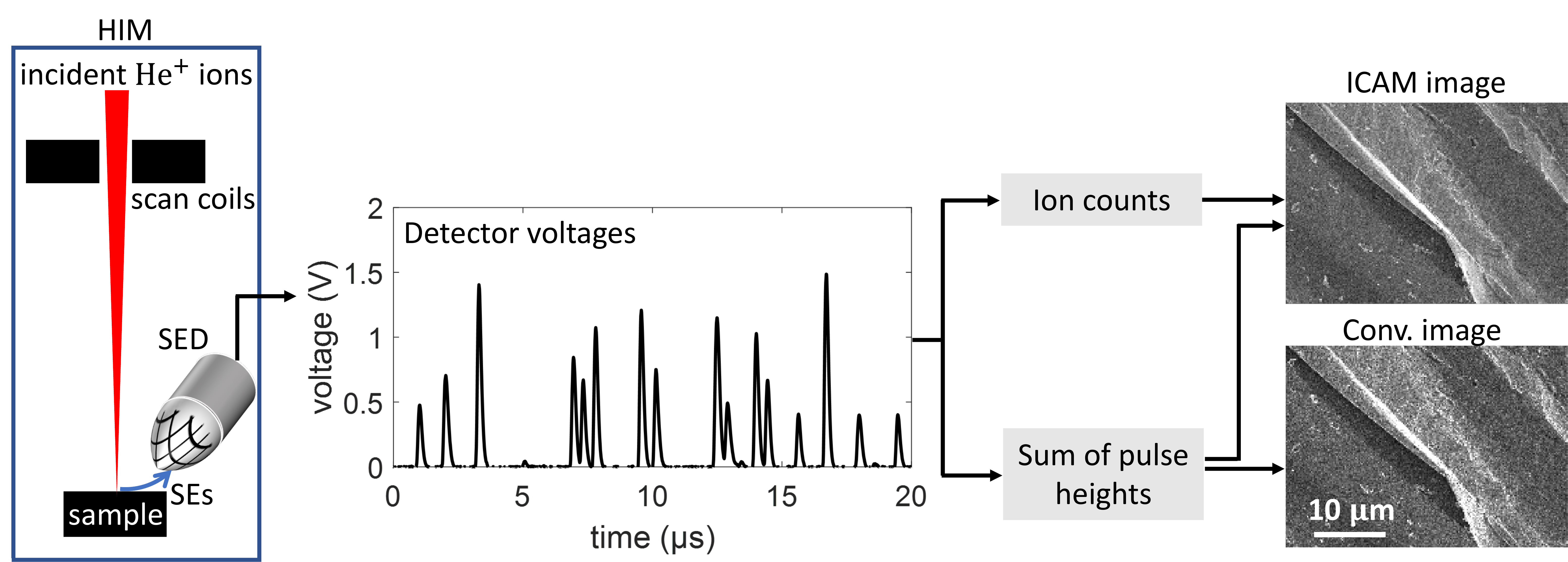}
    \caption{\textbf{Schematic for ion count-aided microscopy in secondary electron imaging}. The signal from the secondary electron detector on a helium ion microscope was outcoupled onto a digitizer for analysis. The heights $\Utilde_i$ and number $\Mtilde$ of the SE voltage pulses in the detector signal were used to create both conventional and ICAM images. ICAM images consistently showed lower noise than conventional images.
    }
    \label{fig:fig1}
\end{figure*}

\Cref{fig:fig1} shows a schematic of our imaging setup. 
We used a Zeiss Orion Plus helium ion microscope operated at 30~$\keV$ beam energy in our experiments.
%To create images, the SED signal needs to be correlated with the position of the ion beam.
We outcoupled both the SED and the beam scan signals into a 12-bit, 100~MHz digitizer (Gage RazorExpress 1642), which sampled both signals at $10~\ns$.
%\VKGcomment{100 MHz is redundant with 10 ns, right?  If you disagree, I would still delete ``bandwidth.''}
\Cref{fig:fig1} shows a $20~\us$ snapshot of the SED voltage signal. 
The SED is typically an Everhart--Thornley detector~\cite{Everhart1960}, which consists of a scintillator followed by a photomultiplier. 
The signal generated by the SED consists of a series of voltage pulses of varying heights, where each pulse corresponds to a burst of detected SEs~\cite{Novak2009a}.
The mean full width at half maximum (FWHM) of the pulses is $\sim160~\ns$~\cite{AGARWAL2021}. 
In our experiments we used beam current $I_b = 0.11~\pA$ (measured with a picoammeter connected to a Faraday cup), which was low enough to make pulse pile-up effects manageable.

To create images, we collected the SED voltage and beam scan signals at pixel dwell times $t_d$ varying between $2~\us$ and $32~\us$ for a 512\,$\times$\,512-pixel image.
These settings correspond to an incident dose $\lambda = \Frac{I_b t_d}{e}$ between 1.4 and 22 ions per pixel,
where $e$ is the elementary charge.
The maximum image size and dwell time were set by the available memory of the digitizer.
Next, using custom MATLAB scripts,
%~\cite{SEImagingCode}
we extracted
the SED voltage pulse heights $\Utilde_i$ and the number of pulses $\Mtilde$ for each pixel.
Neglecting pulse pile-up,
$\Mtilde$ would be the number of incident ions that produced one or more detected SEs.
To account for pile-up, we divide the raw value of $\Mtilde$ by a correction factor that accounts for the probability of pulse overlap to yield a value $\Mtilde_{\textrm{corr}}$ [see Materials and Methods].

\subsection*{Secondary electron yield estimation}
We used the measurements
$(\Mtilde,\Utilde_1,\Utilde_2,\ldots,\Utilde_{\Mtilde})$
to create two pixelwise maps of $\eta$:
\emph{conventional}, which models typical SEI; and
\emph{ion count-aided} (ICA), which implements our source shot noise-mitigated estimator.
While typical SEI uses only $V = \sum_i \Utilde_i$, the performance of ICAM demonstrates the value of using $\Mtilde$ as well.

From its definition as the number of SEs generated per incident particle, it is natural for an
estimate of SE yield $\eta$ to be framed as
\begin{equation}
\label{eqn:eta_intuition}
    \etaHat = \frac{\textrm{estimated number of SEs}}
                   {\textrm{estimated number of incident particles}}.
\end{equation}
In bulk estimates of SE yield, measurements of the beam and sample currents may be used to calculate the numerator and denominator of this expression~\cite{Seiler1983}.
Our nanoscale mapping of $\eta$
uses this intuitive expression with numerator and denominator estimated from
$(\Mtilde,\Utilde_1,\Utilde_2,\ldots$,
$\Utilde_{\Mtilde})$.

\emph{Estimating number of SEs.}
If we knew the mean voltage $\cmu$ produced by the detector in response to one SE,
$\Yhat = \Frac{V}{\cmu}$
would be an unbiased estimator for the number of SEs. 
As detailed in Materials and Methods and in figs.~S1, S2, and S3 in the Supplementary Materials,
we found that a linear probabilistic model accurately describes the statistics of the SE pulses.
In this model, the number of incident particles and the number of emitted SEs both follow Poisson distributions, and the response of the detector to one SE follows a Gaussian distribution.
By fitting this model to the pulse height distributions from samples with different $\eta$, we computed $\cmu = 0.163\,\V$ for the microscope imaging settings used.
We also computed the standard deviation of the detector's voltage response to one SE, $\csig = 0.097\,\V$. 
We used the estimate $\Yhat$ in the numerator of \cref{eqn:eta_intuition} for both the conventional and ICA estimators.

\emph{Estimating number of incident particles.}
Conventionally,
any estimate of the number of incident particles $M$ could depend only on the dose per pixel $\lambda$,
since there is no effort to count incidence events. 
In that case, since the model specifies $M \sim \Poisson{\lambda}$,
the value of $\lambda$ itself is the best estimate of $M$.
In our method, we estimate $M$ using the count of ions that produced at least one SE, $\Mtilde$, corrected for pulse-pileup effects. 
Under our standard statistical assumptions, $\iE{M \smid \Mtilde} = \Mtilde + \lambda e^{-\eta}$.
The $\lambda e^{-\eta}$ term optimally accounts for cases of 0 SE detection and can be a significant fraction of $\Mtilde$ ($\sim16\%$ at $\eta=2$, on average).
Using this expression as an estimate of $M$ lowers the variance of the denominator of \cref{eqn:eta_intuition} by a factor of $e^{-\eta}$
[see Supplementary Materials].

Combining these observations, we can write the expressions for the conventional ($\etaconv$) and ion count-assisted (ICA, $\etaICA$) estimators:
\begin{equation}
    \etaconv = \frac{V/\cmu}{\lambda}
    \label{eqn:etaconv}
\end{equation}
and
\begin{equation}
    \etaICA = \frac{V/\cmu}{\Mtilde_{\textrm{corr}} + \lambda e^{-\etaICA}}.
    \label{eqn:etaTR}
\end{equation}
Note that the second of these is not a formula to evaluate but rather an equation to solve computationally.
Figures S5 and S6 in the Supplementary Materials
show theoretical and experimental calculations of the variance for both $\eta$ estimators, demonstrating the reduction in imaging noise possible with our improved estimate of $M$. 
We also show in Figures S7 and S8
%of the Supplementary Materials
that the images produced by the conventional estimator are equivalent to those produced by the microscope software with suitable scaling.

%\section{Results}
\subsection*{Images with reduced source shot noise}
\begin{figure*}
    \centering
    \includegraphics[width=\linewidth]{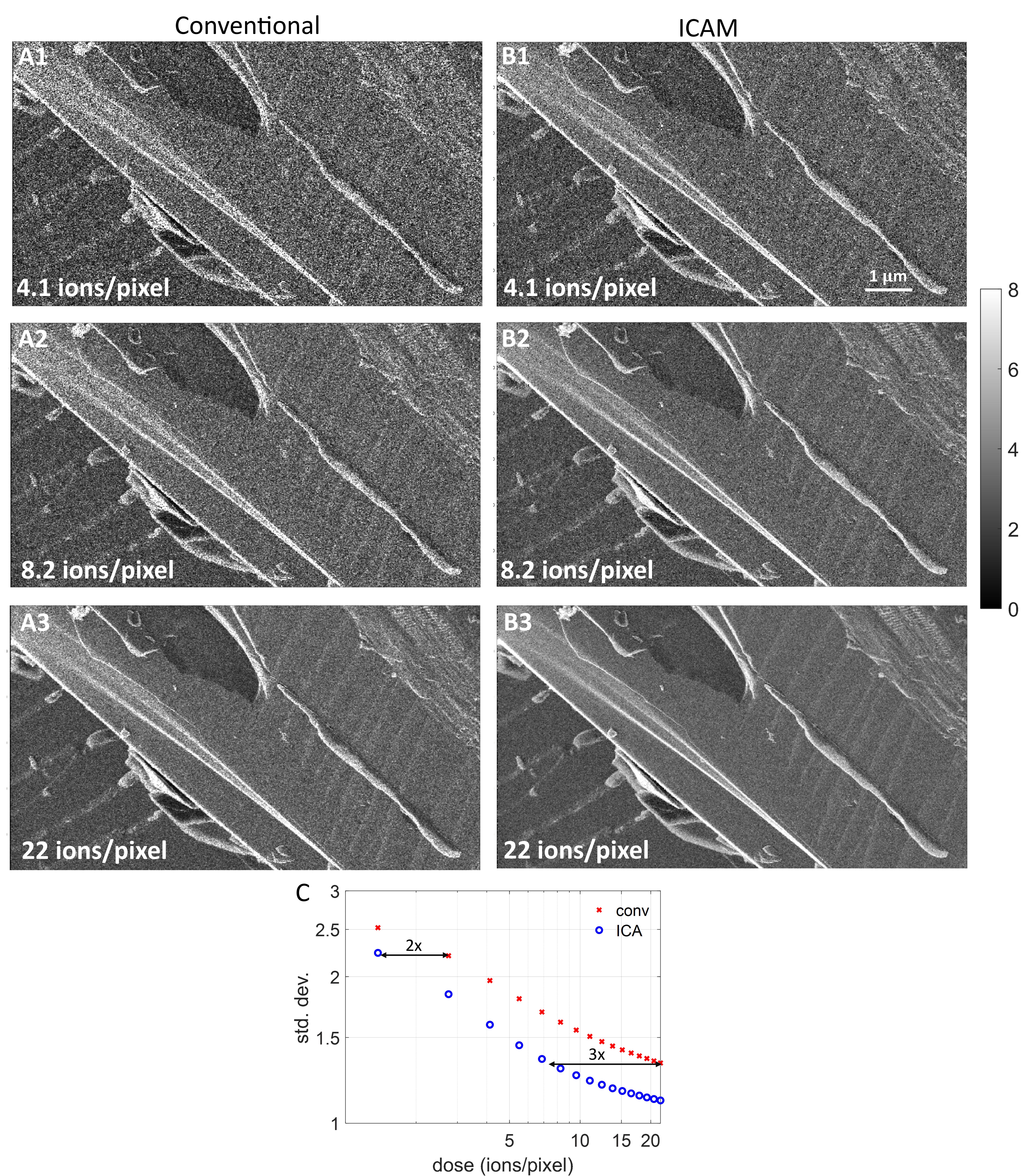}
    \caption{\textbf{ICAM imaging of a silicon sample demonstrating source shot-noise mitigation.}
    \textbf{(A1-A3)} Images created using the conventional SE imaging estimator at doses of 4.1, 8.2, and 22 ions/pixel.
    \textbf{(B1-B3)} Images created using ICAM at the same doses as the conventional images.
    \textbf{(C)} Standard deviation of conventional ({\color{red} $\times$}) and ICAM ({\color{blue} $\circ$}) images as a function of dose. ICAM lowers the dose required to achieve a given standard deviation by a factor between 2 and 3.
    }
    \label{fig:fig2}
\end{figure*}

\Cref{fig:fig2} shows comparisons between conventional and ICAM images of a scratch on a silicon chip at doses of
$\lambda = 4.1$,
$\lambda = 8.2$, and
$\lambda = 22$
ions per pixel.
All images have $288 \times 512$ pixels (cropped from $512 \times 512$ images) and a $10~\um$ horizontal field-of-view. 
Figure S9 in the Supplementary Materials shows uncropped images of this sample at $\lambda = 22$.
These images are SE yield maps;
instead of an arbitrary scale, gray-scale values correspond to physically meaningful values of $\eta$ as indicated by the colorbar.
We can see that for each dose, the ICAM images appear less noisy than the conventional images due to source shot noise mitigation.
The ICAM image at a dose of 8.2 ions/pixel (\Cref{fig:fig2}B2) appears visually similar to the conventional image at a dose of 22 ions/pixel (\Cref{fig:fig2}A3).
\Cref{fig:fig2}C is a plot of the standard deviation measured over all the pixels as a function of the imaging dose for the conventional ({\color{red} $\times$}) and ICAM ({\color{blue} $\circ$}) images.
In addition to the three imaging noise components (source shot noise, target shot noise, and detector noise),
this standard deviation has
a contribution from variations in the features in the sample.
As the dose increases, we expect all three imaging noise components to reduce, so at high doses the standard deviation should asymptotically approach the inherent feature standard deviation.
This is the behavior we observe in \Cref{fig:fig2}C\@.
At low doses, both the conventional and ICAM standard deviations vary inversely with the dose (straight line on a log-log plot), and they show saturation at higher doses.
The horizontal gaps marked in \Cref{fig:fig2}C indicate that ICAM images have the same standard deviations as conventional images at 2- to 3-times the dose,
and the doses selected for \Cref{fig:fig2}A and B illustrate this as well.

Further evidence of the reduction in noise is provided by a calculation of Thong's signal-to-noise ratio (SNR) for both images~\cite{Thong2001}.
This metric aims to calculate SNR for a single SEM image
(without knowledge of ground truth)
by separating the contributions of signal and noise to the image's autocorrelation.
We calculated this SNR to be 5.6 for the ICAM image and 2.8 for the conventional image at $\lambda = 22$.
Figure S11 in the Supplementary Materials
presents a calculation of imaging resolution using Fourier ring correlation~\cite{Saxton1978, vanHeel1982,Saxton1982,vanHeel1987a}
at the $\Frac{1}{3}$ threshold~\cite{Rosenthal2003,Sorzano2017} for both images;
the ICAM image shows a 21\% improvement in resolution.

\begin{figure*}
    \centering
    \includegraphics[width=\linewidth]{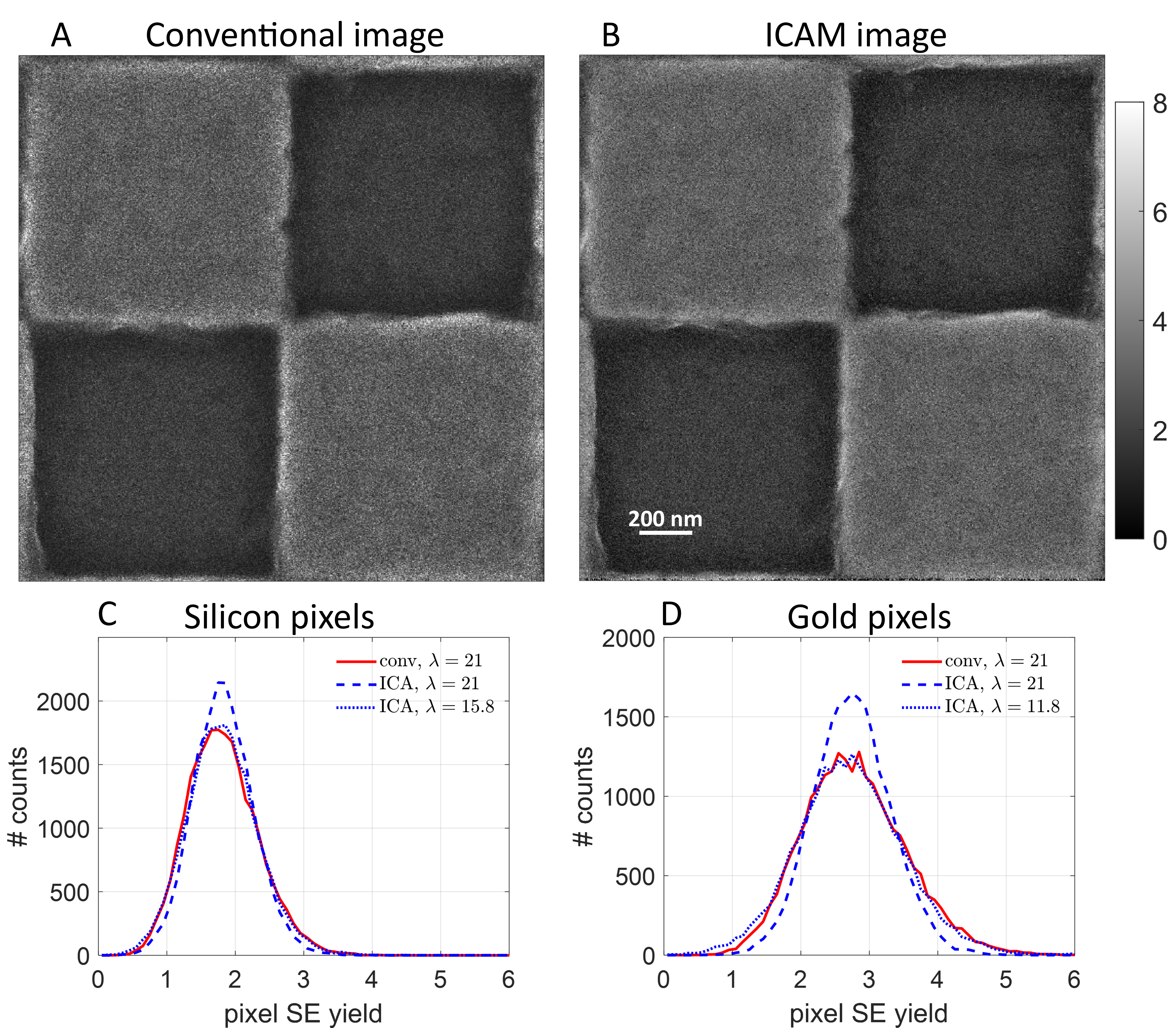}
    \caption{\textbf{ICAM imaging of nanofabricated gold-on-silicon sample.} The total field-of-view is $2~\um$. \textbf{(A)} Conventional image. \textbf{(B)} ICAM image. \textbf{(C)} Histograms of pixel SE yields for darker, silicon pixels for conventional ({\color{red} ---}) and ICAM images ({\color{blue} -\,-\,-}) at the maximum imaging dose of $\lambda = 21$. Since the histograms are computed over pixels with a common composition, their widths indicate imaging noise. ICAM imaging reduces dose requirement by a factor of 1.33, as indicated by the match between ICAM at reduced dose ({\color{blue} $\cdots$}) and the conventional imaging result. \textbf{(D)} Histograms of pixel SE yields for brighter, gold pixels for conventional and ICAM images. ICAM imaging reduces dose requirement by a factor of 1.78.
    }
    \label{fig:fig3}
\end{figure*}

\Cref{fig:fig3}, A and B, shows the conventional and ICAM images of a patterned silicon substrate with $1~\um$ gold squares.
The ICAM image again appears to be smoother and less noisy than the conventional image, especially in the brighter regions that correspond to gold. The lower noise in the ICAM image is also reflected in Thong's SNR:
the ICAM image has an SNR of 2.12, while the conventional image has an SNR of 1.61.

Since this sample has two types of pixels, it provides a good platform for further numerical characterization of the advantages of ICA estimation.
\Cref{fig:fig3}, C and D, are image histograms
with bin width 0.1
for two subsets of pixels in this sample---the darker, silicon pixels, and the brighter, gold pixels---at the same imaging dose.
The histograms plot the frequency of occurrence of different pixel SE yields.
\Cref{fig:fig3}C shows the conventional and ICAM histogram for the dark (silicon) pixels, and
\Cref{fig:fig3}D for the bright (gold) pixels. 
For the silicon pixels, we measured a mean SE yield of 1.82\@. 
We can see that the histogram for the ICAM image is narrower than that for the conventional image at the same dose of $\lambda=21$. 
The width of the ICAM histogram at a lower dose of $\lambda=15.8$ is comparable to that conventional image histogram. 
In other words, comparable image quality was attained using the ICA estimator at a dose that was lower than that for the conventional estimator by a factor of 1.33\@. 
For the gold pixels, we measured a mean SE yield of 2.75\@.
The dose improvement factor in this case was 1.78\@.

Other performance quantifications for the images in \cref{fig:fig2,fig:fig3} also support the conclusion that ICAM improves upon conventional image formation
[see Supplementary Materials].

\subsection*{Theoretical predictions}
\begin{figure*}
    \centering
    \includegraphics[width=\linewidth]{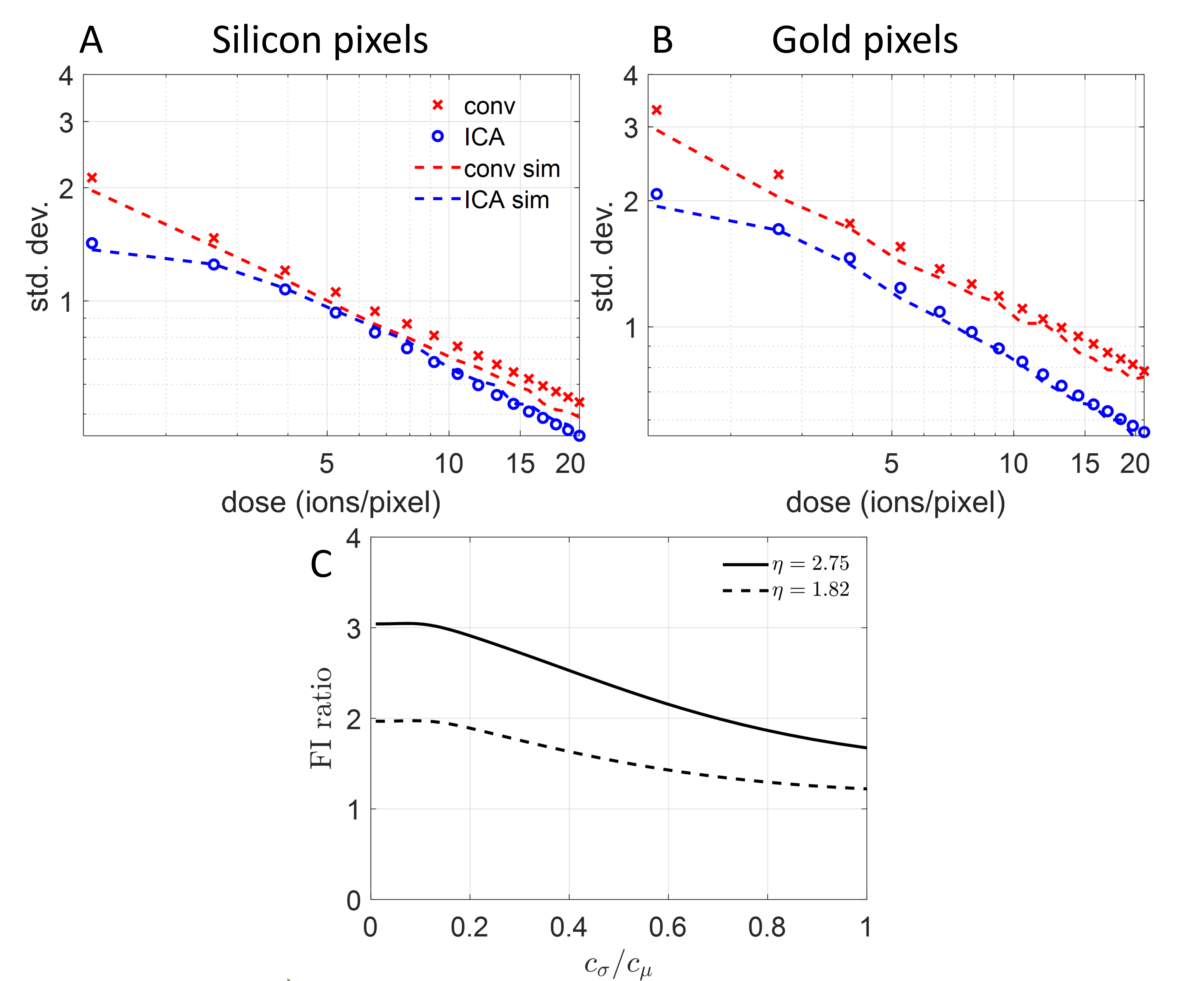}
    \caption{\textbf{Alignment of theoretical predictions and experimental results for noise reduction due to ICA estimation.}
    \textbf{(A)} Standard deviation of conventional and ICAM imaging as a function of imaging dose for $\eta = 1.82$, corresponding to the silicon pixels in \Cref{fig:fig3}. Note that both axes are on a logarithmic scale. The theoretical curves ({\color{red} -\,-\,-} for conventional and {\color{blue} -\,-\,-} for ICAM) are close to the experimental values ({\color{red} $\times$} for conventional and {\color{blue} $\circ$} for ICAM).
    \textbf{(B)} Standard deviation of conventional and ICAM imaging as a function of imaging dose for $\eta = 2.75$, corresponding to gold pixels in \Cref{fig:fig3}.
    \textbf{(C)} Ratio of Fisher Information for conventional and ICAM imaging for $\eta = 2.75$ (---) and $\eta = 1.82$ (-\,-\,-), as a function of $\Frac{\csig}{\cmu}$. For the SED in our HIM, $\Frac{\csig}{\cmu} = 0.6$, and the dose reduction predicted by the FI ratio at this value closely matches our results in \Cref{fig:fig3}.
    }
    \label{fig:fig4}
\end{figure*}

The reduction in imaging noise in \Cref{fig:fig2,fig:fig3} agrees with Monte Carlo performance predictions computed from our model of the imaging process.
%\VKGcomment{Is this detailed in a supplementary section?  It feels like the reader is being told there is a match without being told what the match is to.}
%\VKGcomment{Maybe my issue is that ``theoretical prediction'' is not \emph{wrong}, but we are talking about a Monte Carlo simulation, right?}
\Cref{fig:fig4}, A and~B, compares the theoretical standard deviation as a function of dose with the experimental values we measured for the silicon and gold pixels in \Cref{fig:fig3}, for both conventional and ICA estimators.
%\AAcomment{We can stress again that all 3 sources of noise are going down as dose increases, and we are eliminating one of them.}
%\VKGcomment{I don't really know what you mean by this.}
Unlike the plots in \Cref{fig:fig2}C, we expect no saturation since each square in \Cref{fig:fig3} is almost featureless;
this is exactly what we observe in \Cref{fig:fig4}, A and~B\@.
The experimental standard deviations agree closely with the theoretical values for all doses.
We also see a bigger gap between the standard deviations of conventional and ICAM images at higher SE yield, as expected from the histograms in \Cref{fig:fig3}, C and~D.
As SE yield rises, an increasing fraction of incident particles produce detected SEs, making the estimate of $M$ in \cref{eqn:etaTR} more accurate and improving source shot noise mitigation by the ICAM estimator.
%\AAcomment{Another place to stress on source shot noise reduction.}

The experiments and performance predictions from simulations
%\AAcomment{Do we want to say ``simulations and experiment'' or is that unnecessary?}
are consistent with theoretical analysis through Fisher information (FI).
FI is a measure of the sensitivity of noisy data to the parameter to be estimated;
higher FI indicates proportionately lower imaging noise.
\Cref{fig:fig4}C shows the ratio between the FI from ICAM and conventional measurement at $\lambda = 21$ as a function of $\Frac{\csig}{\cmu}$,
the standard deviation of the contribution of one SE normalized by its mean,
for $\eta = 2.75$ (solid black curve) and $\eta=1.82$ (dashed black curve).
%\AAtext{Remembering that $\cmu$ denotes the mean SED voltage for one detected SE and $\csig$ denotes the standard deviation of this voltage,}
The ratio $\Frac{\csig}{\cmu}$ is a measure of the non-ideality of the SED---the larger this ratio, the more the SED deviates from ideal SE counting.
As $\Frac{\csig}{\cmu}$ approaches zero, the FI ratio is simply $(\eta+1)(1 - \eta e^{-\eta})$, a strictly increasing and approximately linear function of $\eta$~\cite{PengMBG:21}.
%\AAcomment{Should we add a reference to Minxu's TCI paper?}
We can see that the ratio of the FIs of ICAM and conventional measurement is greater than 1 for the entire range of $\Frac{\csig}{\cmu}$, suggesting that ICAM images remain less noisy even for highly non-ideal detectors.

%\VKGcomment{Inserted paragraph break.}

The non-ideality of the detector contributes additional noise to the image, leading to degradation in the FI ratio with increasing $\Frac{\csig}{\cmu}$.
For our system $\Frac{\csig}{\cmu} = 0.6$, and at this value we get FI ratios of about $2.1$ for $\eta=2.75$ and $1.4$ for $\eta=1.82$. 
Due to the additive property of Fisher information, we consequently expect a dose reduction by a factor of $2.1$ at $\eta=2.75$ and $1.4$ at $\eta=1.82$. 
These numbers are close to the experimental dose reductions we obtained for the sample in \Cref{fig:fig3}. 

The SE yield values we report here are not absolute, but are implicitly multiplied by the SED's detection quantum efficiency (DQE)~\cite{Joy1996}. 
We measured the DQE to be about $0.88$ [see Supplementary Materials].
Therefore, all SE yield values quoted here should be divided by this number for absolute SE yields.
For example, the DQE-corrected mean SE yield for the sample in \Cref{fig:fig2} is 3.62, which is in agreement with theoretical predictions of the He-ion SE yield for silicon at $30\,\keV$~\cite{Yamanaka2009}.

Conventional estimates of SE yield require precise measurement of beam current. 
We observed that the beam current value measured by the picoammeter fluctuated by up to 10\% for the same nominal setting of $0.11\,\pA$.
As seen in \cref{eqn:etaconv}, the conventional estimate is inversely proportional to $\lambda$.
Therefore, a 10\% uncertainty in the value of the beam current results in a 10\% uncertainty in $\etaconv$.
In contrast, the ICAM SE yield measurement is much less sensitive to precise knowledge of $\lambda$ because it uses the observed count of ions in addition to knowledge of $\lambda$ to estimate $\eta$ as seen in \cref{eqn:etaTR}.
At $\eta=2.75$, a 10\% variation in $\lambda$ causes 0.8\% variation in $\etaTR$ [see Supplementary Materials].
This reduced sensitivity to precise knowledge of the beam current could be used for removal of stripe artifacts that result from beam current variations~\cite{Seidel2022TCI}.

\section*{Discussion and outlook}
Ion count-aided microscopy differs from other techniques for image denoising and dose reduction in secondary electron imaging. 
Popular methods include image filtering deconvolution~\cite{Roels2016,Vanderlinde2007,Roels2020,Li2016}, adaptive scanning~\cite{Dahmen2016}, and sparse scanning and inpainting~\cite{Kovarik2016,Stevens2018,Pang2021}. 
These methods do not attempt to model the SE generation and detection statistically, and they work as post-processing after conventional SE image generation. 
The ICAM method introduced here is based on statistical modelling of the SE detection process
to improve initial image formation, and it can be combined with various types of post-processing.

Electron count imaging from scintillator-photomultiplier-based detectors has also been demonstrated to increase image SNR and temporal resolution in scanning transmission electron microscopy~\cite{Sang2016,Mittelberger2018,Peters2023}.
In this case, incident electrons that get scattered as they travel through the sample are detected. 
Therefore, at low beam currents, each detected pulse corresponds to one electron, similar to the case of high-energy SEM\@. 
These count-resolved methods are currently limited by pulse pile-up at higher beam currents. 
Since the distribution of heights of piled-up single electron pulses would be identical to that of a single multi-electron pulse, the methods presented here could be applied to model pile-up in these systems to further improve image SNR.

As shown in \Cref{fig:fig3}, the noise-reduction factor due to the ICA estimator increases with increasing SE yield.
%\AAcomment{This sentence is a bit hard to parse since it uses ``reduction in noise...increases'' but I can't think of  way to reframe it to make it clearer. Also, both this and the next paragraph begin with ``As shown in''.}
%\VKGcomment{Is the preceding edit an improvement?}
In fig.~S6B,
we present theoretical calculations of the variance for $\eta$ between 0 and 3.5, and we notice that noise reduction only occurs above an SE yield of about 1\@. 
This limitation would make our method less applicable to SEM at energies above $\sim 2\,\keV$. 
However, these estimators could still be used for low-voltage SEM, where SE yields can be high~\cite{Joy1996b,Joy2006}.
Imaging with other ions, such as neon~\cite{Rahman2011}, would also offer higher SE yields and consequently produce lower-noise images [see Supplementary Materials].
Further, these techniques could also be used for secondary ion imaging and mass spectroscopy with helium and neon beams due to the high sputtering yield~\cite{Wirtz2012}.

The ratio of FI of ICAM and conventional imaging, shown in \Cref{fig:fig4}C, reduces with increasing $\csig$, \textit{i.e.}, increasing variance in the detector's response to each SE single SEs. 
Though the large value of $\Frac{\csig}{\cmu}$ that is typical for Everhart--Thornley detectors limits the performance of ICAM imaging, we were still able to demonstrate a significant improvement. 
Reduction in the variance of the SED's response, by, for example, implementation of solid-state SE detectors~\cite{Ogasawara2016},
would improve both conventional and ICAM imaging while increasing the improvement factor of ICAM.

\subsection*{Acknowledgments}
He Ion Microscope imaging was performed at the Laboratory for Surface Modification (LSM), a center within the School of Arts and Sciences, Rutgers University.
Helpful discussions with Dr.\ Hussein Hijazi and Prof.\ Sylvie Rangan, Rutgers University; and Dr.\ Ben Caplins, National Institute of Standards and Technology, are gratefully acknowledged. 
V.~K. Goyal was supported by the U.S. National Science Foundation grant award number 2039762\@.
X. He was supported by the Boston University Undergraduate Research Opportunities Program.
Ionwerks was supported through an NIH phase II SBIR grant award number 5R44GM106507-03\@.
 
\subsection*{Competing interests}
Schultz/Ionwerks owns patents on secondary electron detector designs and methods of use in correlated SE imaging and ion scattering time of flight measurements.

\subsection*{Data and Materials Availability}
Raw data used to produce all images and graphs herein are available at \\ https://doi.org/10.5281/zenodo.10535916.
%Requests for data and other correspondence should be sent to V. K. Goyal.

\subsection*{Code Availability}
Computer code used to produce all images and graphs is available at \\ https://zenodo.org/doi/10.5281/zenodo.10535998

\newpage

\subsection*{Methods}

\subsubsection*{Experimental setup}
\label{subsec:expt_setup}
We used a Zeiss Orion Plus HIM operating at 30 keV
and a beam current of $0.11\,\pA$ to collect all our data. 
The beam current was set to the lowest stable value to reduce pulse pile-up. 
This beam current corresponds to a dose rate of 0.68 ions per $\us$.
Assuming a focused probe diameter of $1~\si{\nano\meter}^2$ and a pixel dwell time of $32~\si{\micro\second}$ as in Figure 2 of the main paper, we get a dose of $2.2 \times 10^{15}~\textrm{ions}/\si{\centi\meter^2}$.
We outcoupled the voltage signal from the SED into a Gage RazorExpress 1642 CompuScope PCIe Digitizer. 
In preliminary experiments to characterize the response of the SED, we observed that the SE pulses typically had a FWHM of $\sim 160 \, \ns$. 
We decided to use a sampling rate of 100 MS/s (corresponding to a sampling time of 10\,ns) since that results in $\sim20$ samples per pulse, which we considered sufficient to ensure accurate sampling of the pulses' peak voltages.
We typically outcoupled $256 \times 10^6$ samples, corresponding to a total collection time of 2.5\,s. 
This signal was then processed using custom Matlab scripts that counted the pulses and detected their peak voltages.
%~\cite{SEImagingCode}. 

\subsubsection*{Linear Probabilistic Model for SEI}
\label{subsec:ppg_model}
In our model, we denote the (random) number of incident ions at a given pixel by $M$. 
The $i$th ion generates $X_i$ SEs; the total number of SEs generated from the pixel is $Y = \sum_{i=1}^{M}X_i$. 
%Here we assume that all the SEs produced by one incident ion reach the detector simultaneously~\cite{LiMD:19}. 
Both the number of ions $M$ and the $X_i$s are described as Poisson distributed random variables. 
The mean of $M$ is known as the dose $\lambda$.
%and is generally known. 
The mean of $X_i$ is the SE yield $\eta$ and is the quantity we wish to estimate. 
Elementary calculations give $\iE{Y} = \lambda \eta$ and $\var{Y} = \lambda \eta (\eta + 1)$.

Both $M$ and the $X_i$s (or, equivalently, $Y$) are not observed directly in the instrument.
%if they were, we could substitute them into the denominator and numerator of \cref{eqn:eta_intuition} and estimate $\eta$. 
%The generated SEs are detected by the SED, and
We refer to the number of detected SEs as $\Xtilde_i$, 
which is described by a \emph{zero-truncated} Poisson distribution since it is impossible to distinguish between
an incident ion producing zero SEs and there being no incident ion at all.
%tell apart zero detection events from noise.
Finally, $\Xtilde_i$ is mapped to a voltage $\Utilde_i$.
We assume that given $X_i = n$, $\Utilde_i$ is a Gaussian-distributed random variable described by $\normal{nc_{\mu}}{nc_{\sigma}^2}$. 
Here $c_{\mu}$ and $c_{\sigma}^2$ are the mean and variance of the voltage produced by the SED in response to one SE,
with such voltages being independent and additive. An earlier work~\cite{Agarwal2023b} uses the less evocative $(c_1,c_2)$ in place of $(c_\mu,c_\sigma^2)$.
%In prior work, we have used $c_1$ to denote this voltage. $c_{\sigma}$ is the standard deviation in the SED voltage. 
%In prior work, we have used $c_2$ to denote the variance of this voltage.
Overall, the probability density for $\Utilde_i$ is given by
\begin{align}
    f_{\Utilde_i}( u \sMid \eta,c_{\mu},c_{\sigma}^2)
    & = \sum_{j=1}^{\infty} \mathrm{P}_{\Xtilde}(j;\eta)
                            f_Z(u \sMid j,c_{\mu},c_{\sigma}^2) \nonumber \\
    & = \sum_{j=1}^{\infty} \frac{e^{-\eta}}{1-e^{-\eta}}\frac{\eta^j}{j!}
                            f_Z(u \sMid j,c_{\mu},c_{\sigma}^2),
    \label{eq:ppg_pdf}
\end{align}
where $f_Z(u \sMid j,c_{\mu},c_{\sigma}^2)$ is the PDF of a $\normal{jc_{\mu}}{jc_{\sigma}^2}$ random variable.
%\Cref{fig:pbm}E shows an example of the detected output voltage waveform.
%It consists of a series of pulses, each corresponding to detection events from one incident ion. 
%Our task is to estimate $\eta$ given this waveform.
We assume that $\lambda$, $c_\mu$, and $c_\sigma$ are known.

The assumption of linearity in our detection model will hold as long as pulse pile-up is small. 
At small levels of pile-up, the correction term to the raw pulse counts works well. 
At beam currents significantly higher than those used in this study, we expect that more careful consideration will need to be given to the statistics of overlapping pulses, which would lead to improved estimators.

\subsubsection*{Measurement of SED response parameter $c_{\mu}$}
\label{subsec:c1}
To characterize the detector's response to SEs and measure $c_{\mu}$, we imaged a featureless silicon chip to ensure that SE emission was uniform over the scan region. 
Further, we defocused the beam by $5\,\mm$ to reduce beam-induced damage and mitigate variations in SE yield $\eta$ due to local contamination or topography. 
Since $c_{\mu}$ and $c_{\sigma}$ are properties of the detector, we expect them to remain unchanged with variations in the sample; 
i.e., our measurements of $c_{\mu}$ and $c_{\sigma}$ should be constant for samples with variable $\eta$. Therefore, we found it useful to be able to continuously vary $\eta$ to validate our measurements.
This variation of $\eta$ would be difficult to accomplish with physical samples;
instead, we varied the collector bias on the Everhart--Thornley SED to change the effective SE yield at the detector. 
Everhart--Thornley detectors typically have a positively-biased metal cage at the front to attract SEs. 
The default bias on the cage was 500\,V;
we varied this collector bias between 0\,V and 500\,V to get different effective values of $\eta$.

In our model, $c_{\mu}$ is the mean voltage produced by the detector per SE\@. 
Therefore, the most direct method for measuring $c_{\mu}$ would be to have exactly one SE incident on the detector many times and compute the mean voltage produced by the SED. 
Unfortunately, such deterministic irradiation of the SED is not possible, since the production of SEs is a Poisson process. 
One could imagine placing the SED under an electron beam produced by an electron gun, as was done in~\cite{Novak2009a}. 
However, such an experiment requires significant modification of the microscope, which was not possible on our tool.

Instead, we measured $c_{\mu}$ by imaging a sample with low SE yield $\eta$. 
At sufficiently low SE yield, the probability that the incident particle will excite more than one SE is small. 
Therefore, we can assume that most detected pulses are excited by one SE\@. 
Under these conditions, if our model is accurate, we would expect the \emph{pulse height distribution (PHD)}, i.e., the histogram of the peak voltages of the SE pulses, to be approximately Gaussian with a peak at $c_{\mu}$. 
Using this technique, we extracted $c_{\mu}=0.163~\V$ from a histogram measured at $\widehat{\eta} = 0.58$. In the Supplementary Materials, we describe the use of the probabilistic model to fit experimental PHDs at various values of $\widehat{\eta}$.

\subsubsection*{Correcting $\Mtilde$ for pulse pile-up}
As discussed in the paper, ions arriving consecutively in a short time frame can generate overlapping pulses, a phenomenon known as pulse pile-up. 
Consequently, the number of local maxima in a voltage signal can be an underestimation of $\Mtilde$. 
One way to correct for missed detections due to pile-up is by introducing a multiplicative factor:
\begin{equation}
    \Mtildecorr = \frac{\Mtilde}{\gamma_\tau(\Lambda, \eta)},
\end{equation}
where
\begin{equation}
    \gamma_\tau(\Lambda, \eta) = \exp(-\Lambda (1 - e^{-\eta}) \tau)
\end{equation}
is the probability that two pulses arrive within time $\tau$ of each other, and $\Lambda = \Frac{\lambda}{t_d}$ is the dose per unit time. We chose $\tau=0.13~\us$; more details on the choice of $\tau$ can be found in the Supplementary Materials. 

\bibliographystyle{ieeetr}
\bibliography{bibliograph}

\end{document}

\maketitle

{\large
\noindent
{\bf This PDF file includes:} \\[1.2ex]
%\hspace*{5ex} Materials and Methods \\[0.7ex]
\hspace*{5ex} Supplementary text \\[0.7ex]
%\hspace*{5ex} Supplementary references \\[0.7ex]
\hspace*{5ex} Figures S1 to S13 \\[0.7ex]
%\hspace*{5ex} Supplementary reference 50\\[0.7ex]
%\hspace*{5ex} References
}

%\tableofcontents

\newpage

\section*{Supplementary Text}
\noindent
{\bf \emph{Contents}}
\begin{itemize}
\setlength{\itemsep}{0pt}
    \item \Cref{subsec:expt_setup}: Distribution of pulse interarrival times.
    \item \Cref{subsec:phd_fitting}: Fitting pulse height distributions using $\eta$ estimates.
    \item \Cref{subsec:estimators}: Design and analysis of estimators for ion count $M$ and SE yield $\eta$.
    \item \Cref{subsec:conv_software}: Comparison between our conventional images and images from instrument software.
    \item \Cref{subsec:SNR_FRC}: Further quantitative analysis through signal-to-noise ratio and resolution.
    \item \Cref{subsec:DQE}: Computation of detector quantum efficiency.
    \item \Cref{subsec:current_sensitivity}: Analysis of sensitivity of estimators to beam current fluctuation.
    \item \Cref{subsec:heavier_ions}: Analysis of relative error dependence on scaling of SE yield.
\end{itemize}

\section{Distribution of pulse interarrival times}
\label{subsec:expt_setup}

\begin{figure}[b]
  \begin{center}
    \begin{tabular}{@{}c@{\,}c@{\,}c@{}}
      \includegraphics[width=0.36\linewidth]{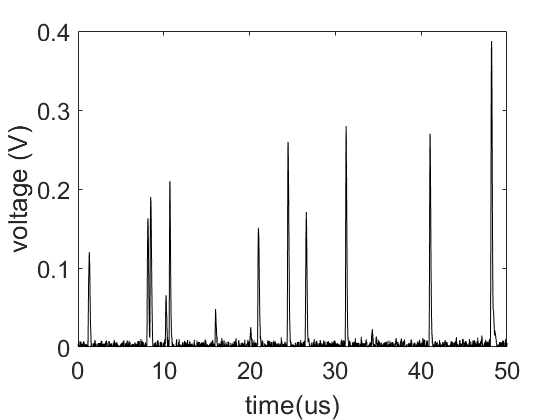} &
      \includegraphics[width=0.36\linewidth]{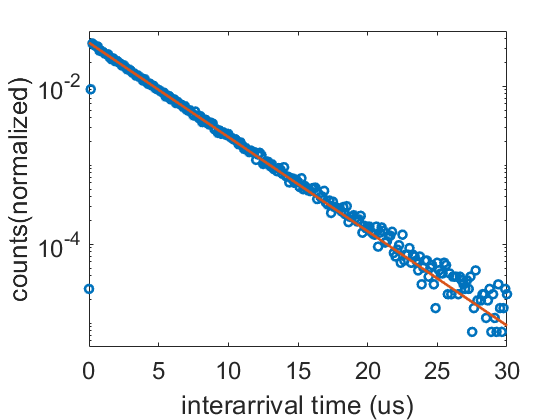} \\
      {\small (A) SED signal} &
      {\small (B) Inter-arrival time distribution}
    \end{tabular}
  \end{center}
    \caption{Initial characterization of SED signal.
    The interarrival time distribution follows an exponential distribution and allows estimation of $\eta$.}
    \label{fig:signal_charaterization}
\end{figure}

\Cref{fig:signal_charaterization}A shows a 50~\si{\micro\second} example of the outcoupled voltage signal. 
As preliminary evidence for our probabilistic model, \Cref{fig:signal_charaterization}B shows a histogram of the interarrival times between successive pulses for the same dataset as \Cref{fig:signal_charaterization}A\@. 
This data was obtained from the silicon sample by setting the SED collector bias to 0\,V\@. 
If $\Mtilde$ follows a Poisson distribution as discussed in \Cref{subsec:mtilde},
the interarrival times should be exponentially distributed with parameter $\lambda(1-e^{-\eta})$. 
As indicated by the linearity of the interarrival time distribution on the semilog plot in \Cref{fig:signal_charaterization}B,
the interarrival times indeed follow an exponential distribution. 
The parameter of the distribution was obtained from a linear fit (solid orange line) to be $0.275$ per ${\si{\micro\second}}$. 
With $\lambda = 0.625$, we use \cref{eqn:Mtilde} to get $\widehat\eta_{\Mtilde} =0.58$. 
Such estimates provide a useful benchmark for the characterization of other ICA estimators, as detailed in \Cref{subsec:etaexpt}.

%\AAcomment{Should the linear model come here, before the measurement of cmu? It seems a bit weird to present fits to the model without first stating the model. I think it would be appropriate to have it in `methods'.}

\section{Fitting pulse-height distributions using $\eta$ estimates}
\label{subsec:phd_fitting}
We fitted the distribution of pulse heights at different values of $\eta$ to the probabilistic model described in Methods. 
This fitting yields the parameter $c_{\sigma}$, the variance of the SED's response to one SE.
While $c_{\sigma}$ does not enter into estimate computations, it is critical in performance predictions.

\Cref{fig:phd_fit1,fig:phd_fit2} show the experimental PHD ({\color{blue} $\circ$}) and the voltage probability distribution from our model ({\color{orange} ---}) for various values of $\eta$. 
Recall that the value of $\eta$ was varied by changing the SED collector bias (from $0\,\V$ for the lowest $\eta$ to $500\,\V$ for the highest);
the physical sample is silicon in all cases. 
For each value of $\eta$, the left panel shows the PHD and model fit on a linear scale, and the right panel shows the same fits on a log scale. 
The figure captions also indicate the ICA estimate of $\eta$ used for these fits.
We note that in obtaining these fits, we added a zero-mean Gaussian component to fit the low-voltage noise in the PHD\@.
For all fits, this component had a relative amplitude between 0.76 and 0.78 compared to the signal, and a constant standard deviation of 0.02\@. 
We see that for all values of $\eta$, the PHD is very well described by the model with $c_{\mu} = 0.163\,\V$ and $c_{\sigma} = 0.097\,\V$.
For low values of $\eta$ (\Cref{fig:phd_fit1}), the entire PHD is fit by the model. 
At higher values of $\eta$ (\Cref{fig:phd_fit2}), a high-voltage feature emerges in the PHD around $1.6\,\V$, which causes deviations between the PHD and the model fit. 
This feature can be seen most clearly in the log plots. 
We expect that this feature occurs because of pulse height saturation. 
The maximum voltage the SED produced in all of our experiments was $1.8\,\V$, which corresponds to about 11 SEs.
As $\eta$ increases, the probability of producing more than 11 SEs increases, and these pulses get saturated at or near $1.8\,\V$ causing the feature seen in the PHD\@. 
Inclusion of such saturation in our model would improve the fits at higher values of $\eta$.

\newlength{\phdfit}
\setlength{\phdfit}{0.49\linewidth}
\begin{figure}
  \begin{center}
    \begin{tabular}{@{}c@{\,}c@{\,}}
      \includegraphics[width=\phdfit]{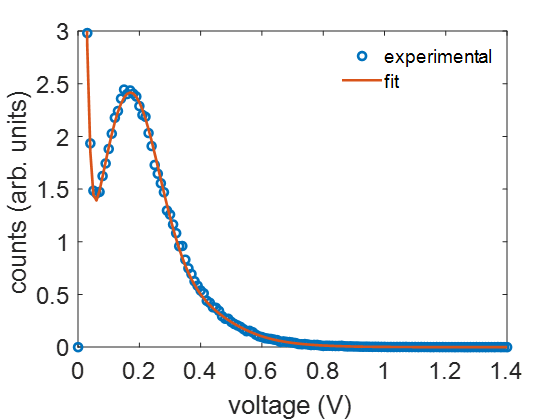} &
      \includegraphics[width=\phdfit]{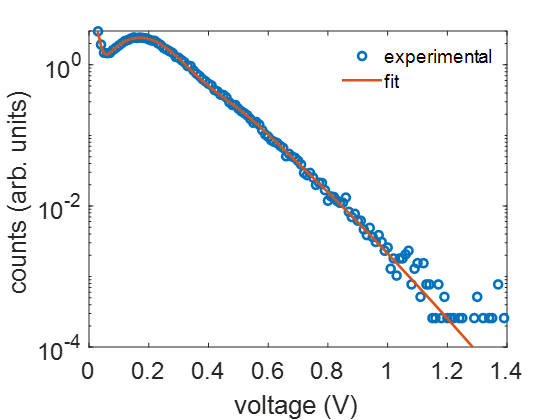} \\
      {\small (A) Linear, $\etaICA = 0.58$} &
      {\small (B) Log, $\etaICA = 0.58$} \\
      \includegraphics[width=\phdfit]{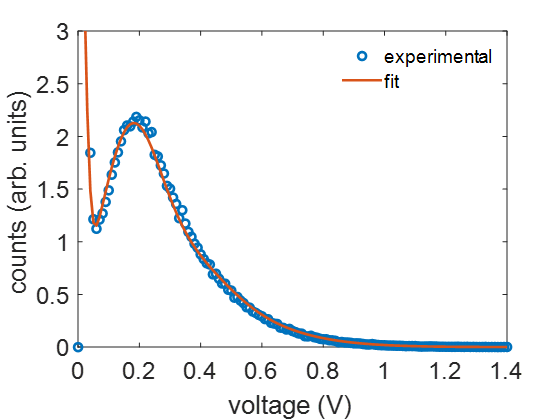} &
      \includegraphics[width=\phdfit]{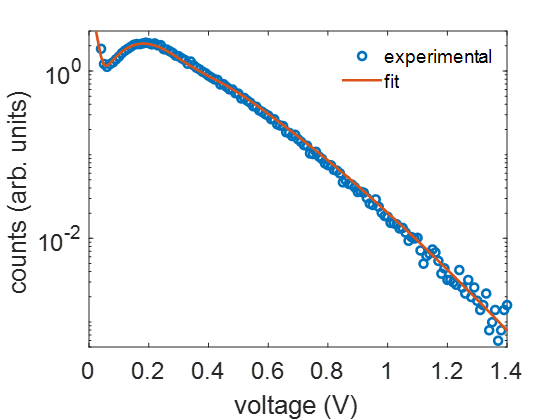} \\
      {\small (C) Linear, $\etaICA = 1.1$} &
      {\small (D) Log, $\etaICA = 1.1$} \\
    \end{tabular}
  \end{center}
    \caption{Pulse height distribution (blue circles) and model fit (solid orange line) for various values of $\eta$ on linear (left) and log (right) scale. Caption indicates the ICA and fit $\eta$ values. In all cases, the fits used $c_{\mu} = 0.163~V$ ad $c_{\sigma} = 0.097~V$.}
    \label{fig:phd_fit1}
\end{figure}

\begin{figure}
  \begin{center}
    \begin{tabular}{@{}c@{\,}c@{\,}}
      \includegraphics[width=\phdfit]{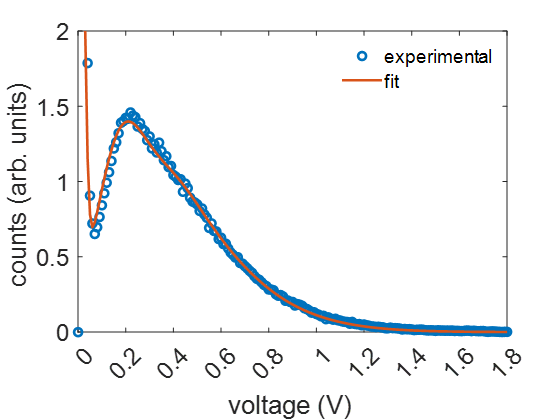} &
      \includegraphics[width=\phdfit]{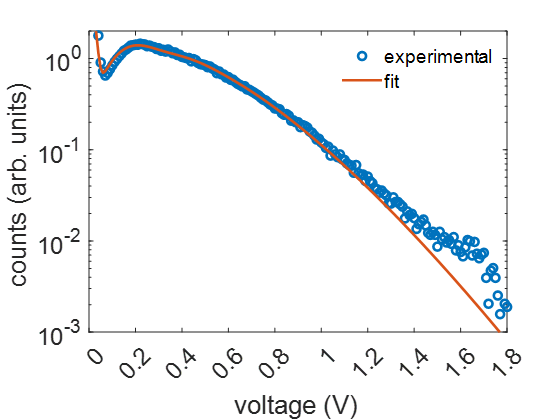} \\
      {\small (A) Linear, $\etaICA = 2.1$} &
      {\small (B) Log, $\etaICA = 2.1$} \\
      \includegraphics[width=\phdfit]{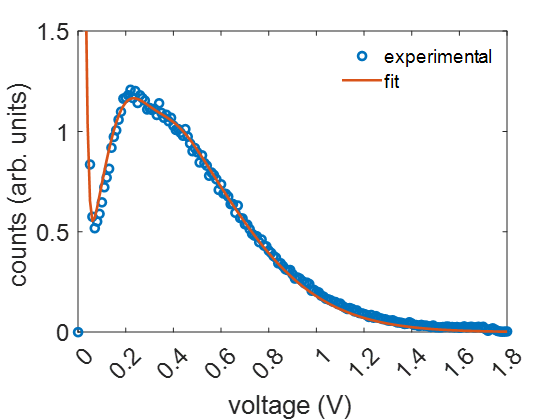} &
      \includegraphics[width=\phdfit]{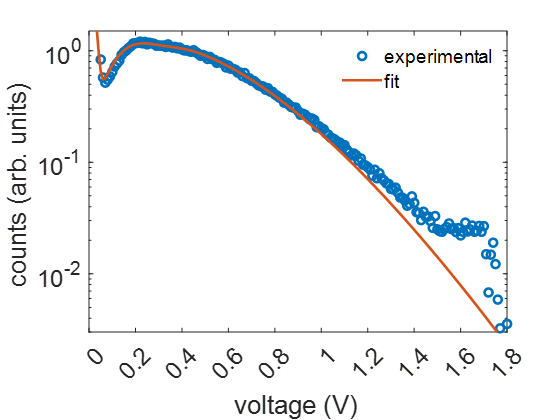} \\
      {\small (C) Linear, $\etaICA = 2.57$} &
      {\small (D) Log, $\etaICA = 2.57$} \\
      \includegraphics[width=\phdfit]{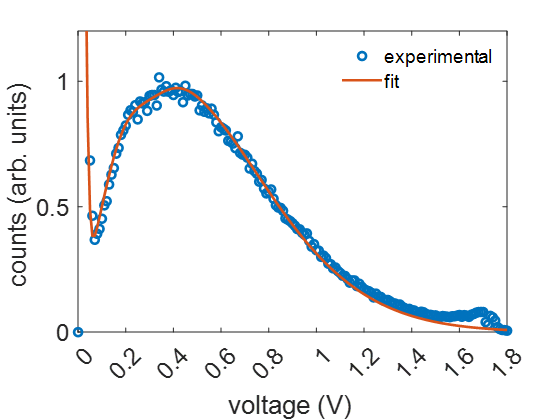} &
      \includegraphics[width=\phdfit]{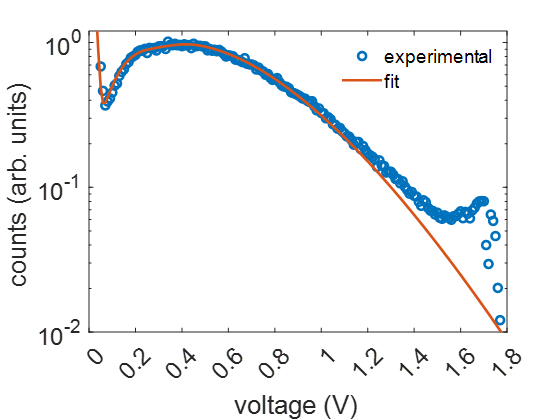} \\
      {\small (E) Linear, $\etaICA = 3.25$} &
      {\small (F) Log, $\etaICA = 3.25$}
    \end{tabular}
  \end{center}
    \caption{Same as \Cref{fig:phd_fit1} for higher values of $\eta$. Deviation between model and experimental PHD can be seen at higher voltages due to pulse saturation.}
    \label{fig:phd_fit2}
\end{figure}

%\clearpage

%\newpage
%\section*{Supplementary Text}

%This document provides supporting information for the main paper.
%In \Cref{sec:theory}, we describe the conventional and time-resolved (TR) estimators in detail and builds intuition for their functional forms. 
%In \Cref{sec:experiment}, we characterize the secondary electron detector (SED) to extract the model parameters $c_{\mu}$ and $c_{\sigma}$, and use these parameter values to estimate $\eta$ for bulk, featureless samples as a precursor to the pixelwise $\eta$ mapping presented in the main paper. 
%Next, we compare the noise in experimental estimates of $\eta$ to theoretical estimates obtained from Monte Carlo simulations~\cite{Agarwal2023b}. 
%We also establish the close similarity between the software image and the conventional estimator image to justify using the latter to represent typical performance. 
%Further, we characterize the SNR and resolution of the TR image compared to the conventional image, and we present a measurement of the SED's detective quantum efficiency (DQE).

%\section{Theoretical description of estimators}
%\label{sec:theory}

\section{Estimators for Ion Count $M$ and SE Yield $\eta$}
\label{subsec:estimators}
%In this section we will build intuition about conventional and time-resolved estimators for the SE yield. 
As discussed in the paper, an intuitive form for SE yield estimators is
\begin{equation}
    \widehat{\eta} = \frac{\textrm{number of SEs detected}}{\textrm{number of incident particles}}.
    \label{eqn:eta_intuition}
\end{equation}
In this section, we introduce and analyze several estimators of this form.
They differ in how the ion count $M$ is estimated for use in the denominator of \cref{eqn:eta_intuition}.
Since some estimators for $M$ have dependence on $\eta$, which is unknown, the substitution can yield an equation to solve for $\etaHat$.
%First, we will describe our abstract model for SE generation and detection in SEI\@. 
%Then, we will describe conventional and time-resolved estimators for the SE yield $\eta$.

%\AAcomment{Should the phrase `time-resolved' be used at all?}

\subsection{Oracle-aided estimator}
\label{subsec:oracle}
If $Y$ and $M$ were provided by an oracle, with $M > 0$, $\eta$ could be estimated as
\begin{equation}
    \etaOracle = \frac{Y}{M}.
    \label{eqn:oracle}
\end{equation}
Conditioned on $M$ being greater than 0, this is an unbiased and efficient estimate of $\eta$.
Though not implementable from measurements we have available,
%Since both $Y$ and $M$ are not directly measured, this is not a practical $\eta$ estimator. 
it will serve as a useful theoretical bound on the performance of other, more practical, estimators.

\subsection{Conventional estimator}
\label{subsec:conventional_pbm}
%This estimator is our first practical attempt to directly estimate $\eta$. 
%Since our goal is $\eta$ estimation, we have to figure out how to estimate the numerator and denominator of \Cref{eqn:eta_intuition}. 

For estimating $Y$, suppose the sum of the peak voltages from all the pulses generated by the ions incident on the pixel within the dwell time is $V = \sum \Utilde_i$. 
Then, $V\sim \normal{c_{\mu}Y}{c_{\sigma}^2 Y}$.
Thus, $V/c_\mu$ is the maximum likelihood (ML) estimator for $Y$, and
it has conditional variance $c_\sigma^2 y / c_\mu^2$ given $Y = y$.
%therefore, for the numerator of \cref{eqn:eta_intuition}. 
%We will describe how we estimated $c_{\mu}$ in \Cref{sec:experiment}.
As discussed in the paper, $\lambda$ is an unbiased estimate of $M$.
%For estimating the denominator, since we know that $M$ is Poisson distributed with mean $\lambda$, and $\lambda$ can be easily computed from the beam current and pixel dwell time, we could use $\lambda$ as an estimator for $M$. 
%This estimator will be biased in general, since $M$ is random and will likely be different from $\lambda$. 
%It will have 0 variance, since $\lambda$ is constant.
Therefore, the variance (and mean-squared error (MSE)) of this estimate is given by
\begin{equation}
    %\textrm{MSE}_{\hat{M}} =
    \E{(M-\lambda)^2} = \lambda.
    \label{eqn:MSE_lambda}
\end{equation}
Substituting these estimates into \cref{eqn:eta_intuition}, we arrive at the \emph{conventional} estimator:
\begin{equation}
    \etaconv = \frac{V/c_{\mu}}{\lambda}.
    \label{eqn:conv}
\end{equation}
It is unbiased and has MSE
\begin{equation}
\label{eq:conventional_MSE}
    \E{\left( \eta - \etaconv \right)^2} = \frac{\eta(\eta + 1 + c_\sigma^2/c_\mu^2)}{\lambda}.
\end{equation}
This MSE expression shows the contribution from SED noise ($c_\sigma > 0$).
Having quadratic (rather than linear) dependence on $\eta$ is a manifestation of source shot noise~\cite{PengMBG:21}.

We note that the form of this estimator differs from the way SE imaging is conventionally performed. 
Typically, SEI does not aim to directly estimate $\eta$; instead, the aim is to create an image that qualitatively represents variations in $\eta$ over the sample. 
Therefore, \cref{eqn:eta_intuition} is not directly involved in conventional imaging. 
Instead, the voltage waveform is sampled at a constant rate.
Each of the detected pulses has a FWHM of $\sim 160\,\ns$, and the sampling time needs to be lower than this width to ensure that each pulse is accurately sampled. 
In practice, a sampling time of $100\,\ns$ is typically used. 
This sampled voltage is quantized to (usually) 8 bits by an ADC\@. 
Other unspecified operations may also be applied to the SED voltage signal before quantization. 
The final image is a map of this `pixel brightness'.

We found that our conventional estimator and `true' conventional imaging had almost identical performance in terms of imaging noise, after the true conventional image was appropriately scaled. 
In \Cref{subsec:conv_software}, we show images created by both methods; their appearances and histograms are close to identical.
Therefore, it is appropriate to use the conventional estimator to represent the performance of existing SEI, as done in the main paper.
%Therefore, from now on, we will use `conventional' to mean both the conventional and improved conventional estimators. We keep in mind that this conventional estimator is a `steelman' version of true conventional PBM, since neither $c_{\mu}$ is nor the peak pulse voltages are measured conventionally.

\subsection{Quotient-mode estimator}
\label{subsec:qm}
In the previous section, we described how $V/c_{\mu}$ is the ML estimator for $Y$. 
We also found the variance for using $\lambda$ as an estimate of $M$ to be equal to $\lambda$. 
To improve the estimation of $M$, we notice that in addition to the peak voltage of each pulse, we also know the total number of detected pulses, $\Mtilde$.
Since each pulse corresponds to events from one incident ion, we can use $\Mtilde$ as an estimate of $M$.
We note that this estimator will be biased, since it does not account for the cases where an incident ion produces 0 SEs, and therefore it underestimates $M$.  
The MSE for this estimator is given by
\begin{equation}
    %\textrm{MSE}_{\hat{M}} =
    \E{(M-\Mtilde)^2} = \lambda e^{-\eta}(1 + \lambda e^{-\eta}).
    \label{eqn:MSE_mtilde}
\end{equation}
For large $\eta$, this expression becomes very small, indicating that the MSE from this estimate will be lower than that in \cref{eqn:MSE_lambda}. 
However, for small $\eta$, this MSE is approximately $\lambda(1+\lambda)$, which is larger than that in \cref{eqn:MSE_lambda}.

Using $\Mtilde$ as an estimator for $M$, we get the \emph{quotient-mode} (QM) estimator:
\begin{equation}
    \etaQM = \frac{V/c_{\mu}}{\Mtilde}.
    \label{eqn:qm}
\end{equation}
The accuracy of $\Mtilde$ as an estimate of $M$ given in \cref{eqn:MSE_mtilde} is reflected in the MSE of $\etaQM$.
At low $\eta$, many incident ions produce 0 SEs, and therefore $\Mtilde$ significantly underestimates $M$ and results in high MSE.
At high $\eta$, almost all incident ions produce at least one detected SE, and therefore $\Mtilde$ is an accurate estimate of $M$ and results in low MSE\@.
An abandoned patent application~\cite{Zeiss_QM_patent} contains a description similar to the quotient-mode estimator,
though this was apparently not implemented in any product.
This may be due to its poor performance at low $\eta$.

%The name of this estimator is inspired from a presentation made by Dr. John Notte from Zeiss \AAcomment{Is there a citation for this?}.
%Intuitively, we expect $\Mtilde$ to be smaller than $M$, because we will only get a pulse when the incident ion induces at least one detected SE\@. 
%Therefore, $\Mtilde$ does not include ions that result in 0 SEs emitted. 
%At high values of $\eta$, this will not be a big problem, since almost all ions will emit at least one SE\@. 
%However, at low values of $\eta$, $\widetilde{M}$ will differ significantly from $M$. 
%This bias in $M$ estimation also makes \Cref{eqn:qm} a biased estimator of $\eta$.

\subsection{Maximum likelihood-inspired estimator}
\label{subsec:ML}
Inspired by the QM estimator, we could try to improve upon $\Mtilde$ as an estimate of $M$. 
Such an estimator should depend on $\Mtilde$, but must also somehow include a correction for the cases in which 0 SEs are detected.
Since the mean number of incident ions is $\lambda$, and the number of SEs emitted per ion has a Poisson distribution with mean $\eta$, the expected number of ions that result in 0 SEs is $\lambda e^{-\eta}$. 
Therefore, $\Mtilde + \lambda e^{-\eta}$ should be an improved, unbiased estimator for $M$. 
Indeed, through more rigorous arguments detailed in~\cite{Agarwal2023b}, this quantity is the minimum MSE estimator for $M$.
The MSE for this estimator can be calculated as
\begin{equation}
    %\textrm{MSE}_{\hat{M}} =
    \E{(M-(\Mtilde+\lambda e^{-\eta}))^2} = \lambda e^{-\eta},
    \label{eqn:MSE_mli}
\end{equation}
which is uniformly lower than the expression in \cref{eqn:MSE_lambda} by a factor of $e^{-\eta}$.
Using this estimate of $M$ in \cref{eqn:eta_intuition}, we get 
\begin{equation}
    \etaMLI = \frac{V/c_{\mu}}{\Mtilde + \lambda e^{-\etaMLI}}.
    \label{eqn:mli}
\end{equation}
This equation can be solved with a suitable root-finding algorithm.
We refer to this estimator as \emph{ML-inspired} (MLI), rather than ML,
because a true ML estimator would require maximization of the likelihood for the joint observation $(V,\Mtilde)$ under the probabilistic model described in Methods. 
Such an estimator cannot be expressed using a simple analytical expression. 
Because we are using efficient estimators for both $Y$ and $M$,
we expect the performance of $\etaMLI$ be close to a true ML estimator,
%\AAcomment{Is there a way to make this statement more technically accurate? I'm trying to qualitatively justify why this estimator's performance should be close to the true ML estimator.}
and we confirmed through Monte Carlo simulations that this is indeed the case.
This estimator is also more computationally tractable than a true ML estimator for large datasets.

%\label{subsubsec:gamma}
\textbf{Correcting the MLI estimator for pulse pile-up}: 
As described in Methods, we accounted for pulse pile-up by introducing $\Mtildecorr$:
\begin{equation}
    \Mtildecorr = \frac{\Mtilde}{\gamma_\tau(\Lambda, \eta)},
\end{equation}
where
\begin{equation}
    \gamma_\tau(\Lambda, \eta) = \exp(-\Lambda (1 - e^{-\eta}) \tau)
\end{equation}
is the probability that two pulses arrive within time $\tau$ of each other, and $\Lambda = \Frac{\lambda}{t_d}$ is the dose per unit time~\cite{Agarwal2023b}.
Substituting $\Mtilde$ by $\Mtildecorr$ in \cref{eqn:mli} yields the \emph{count-corrected ML-inspired} estimator:
\begin{equation}
    \etaCCMLI = \frac{V/c_{\mu}}{\Mtilde / \gamma_\tau(\Lambda, \etaCCMLI) + \lambda e^{-\etaCCMLI}}.
    \label{eqn:gamma}
\end{equation}

\begin{figure}
  \begin{center}
    \begin{tabular}{@{}c@{\,}c@{\,}}
      \includegraphics[width=0.48\linewidth]{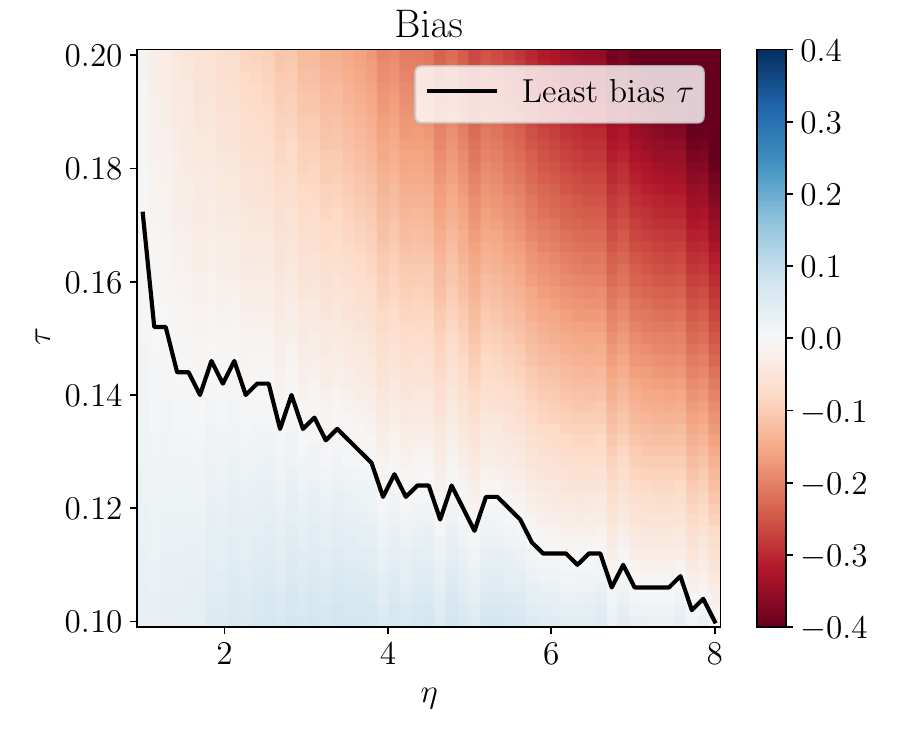} &
      \includegraphics[width=0.48\linewidth]{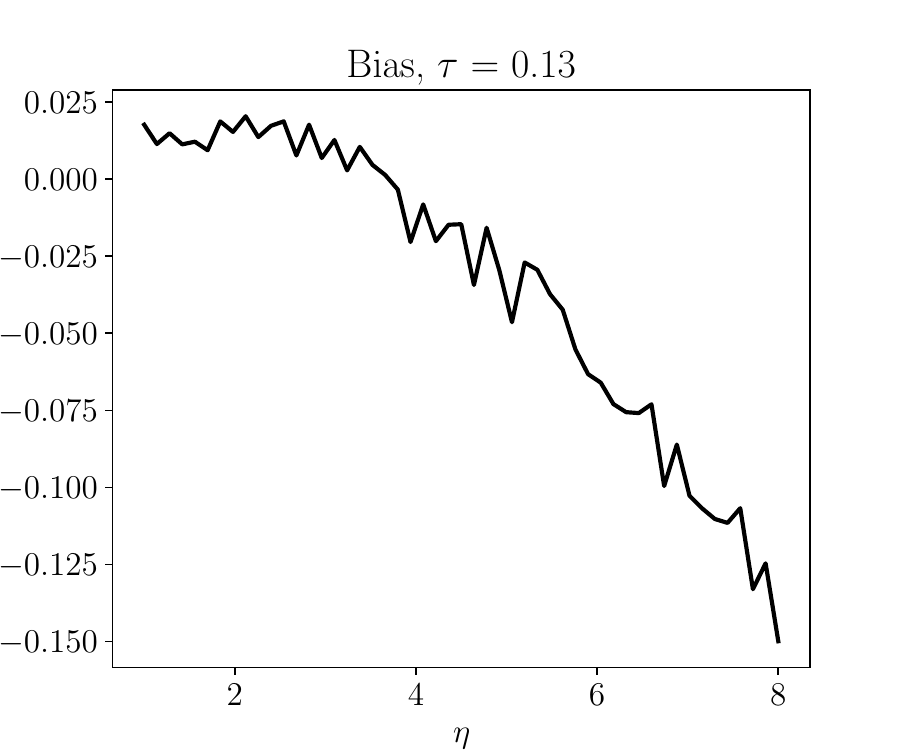} \\
      {\small (A) Bias at different $\eta$ and $\tau$} &
      {\small (B) Bias at $\tau = 0.13$}
    \end{tabular}
  \end{center}
    \caption{Bias of the count-corrected ML-inspired estimator $\etaCCMLI$, defined in \cref{eqn:gamma}, obtained through Monte Carlo simulation with 10\,000 trials.
    The rate $\Lambda$ is 0.625 ion/$\si{\micro\second}$. The line in (A) is $\tau$ at which the absolute bias is minimized at each $\eta$.}
    \label{fig:gamma}
\end{figure}

One remaining challenge is choosing an appropriate $\tau$. 
In addition to the inter-arrival time, both widths and heights of the pulses determine whether a given set of pulses overlap and merge into one large pulse. The pulses have randomly varying widths and heights, the latter of which depends on the number of SEs generated, which further depends on unknown $\eta$. 
Nevertheless, as shown in~\Cref{fig:gamma}, choosing $\tau$ around 0.12\,$\si{\micro\second}$ to 0.14\,$\si{\micro\second}$ yields an estimator $\etaCCMLI$ with a small bias for
$\eta \in [1, 8]$, a suitable range of interest for HIM\@. 
These plots were computed using a Monte Carlo simulation whose details are described in \Cref{subsec:etaexpt}.
%\VKGcomment{Is that true?} \AAcomment{I added a couple more sentences to the description of the simulation in section 2.6.}
For all the results in the paper, we chose $\tau = 0.13\,\si{\micro\second}$. 
\Cref{fig:gamma}B shows the bias in the CCMLI estimator for this choice of $\tau$;
in absolute value, it remains below 0.02 for the range of $\eta$ imaged in the paper.

The count-corrected MLI estimator was used to compute all the ICAM images in the paper.
%\AAcomment{Is this too confusing? Should I just call it the ICA estimator and not use MLI?}
In the paper, the estimator is referred to simply as the ion count-aided estimator $\etaICA$ because we do not wish to compare different ion count-aided estimators there
(one could argue that $\etaQM$ and $\etaMLI$ without count correction are also ion count-aided estimators because they use knowledge of $\Mtilde$).
In the subsequent sections of this document, we return to the use of $\etaICA$ to refer to this estimate, which incorporates our modeling insights without introducing high computational complexity through full use of the probabilistic model or the possibility of an $\eta$-dependent $\tau$ value.
%\VKGcomment{Your point about not straying too much from the use of ICA is well taken.  With that statement, we could change the notation in S2.6 and Figure S6?}

\subsection{Pulse-count estimator}
\label{subsec:mtilde}
All the practically implementable $\eta$ estimators described so far require knowledge of $c_{\mu}$, the mean voltage produced by 1 detected SE\@. 
%We will describe our measurement of $c_{\mu}$ in \Cref{subsec:c1}. 
However, it would be good to validate the $\eta$ estimates from these estimators against an independent measurement of $\eta$ that does not rely on knowledge of $c_{\mu}$. 
We could potentially measure $\eta$ indirectly from the sample current from the silicon sample,
but such a measurement cannot be expected to agree with $\eta$ measured at the SED due to non-ideal detection quantum efficiency (DQE)~\cite{Joy2008,AGARWAL2021}.

Instead, we will use just the count of SE pulses $\Mtilde$ to estimate $\eta$. 
Since $M$ is modeled as $\Poisson{\lambda}$ and the count is thinned by ion incidences that produce no detected SEs,
%By the same arguments as in \Cref{subsec:ML}, we can conclude that $\Mtilde$ should have a Poisson distribution with a mean lower than $\lambda$. 
%The reduction should be exactly equal to the expected number of ions that produce no SEs. 
%Therefore,
$\Mtilde \sim \Poisson{\lambda(1-e^{-\eta})}$. 
If we observe $\Mtilde$ pulses, we can estimate $\eta$ as
\begin{equation}
    \widehat\eta_{\Mtilde} = -\log\left(1-\frac{\Mtilde}{\lambda}\right).
    \label{eqn:Mtilde}
\end{equation}
The properties of this estimator are discussed in~\cite{Agarwal2023b}. 
Briefly, we expect this estimator to be accurate at low $\eta$, where it is very likely to have $\Mtilde < \lambda$. 
At high $\eta$, almost all the incident ions will produce at least one SE\@. 
Therefore, $\Mtilde$ will be pretty close to $\lambda$ and might even exceed it if $M>\lambda$ in some experiments. 
The estimate in \cref{eqn:Mtilde} is likely to be highly inaccurate in this case. 
However, at low $\eta$, the estimator will be accurate, and we will use it to benchmark the results of other estimators in \Cref{subsec:etaexpt}.

%\section{Theoretical performance of estimators}
%\label{subsec:eta_performance_theoretical}
%\Cref{fig:ppg-theoretical} shows the mean squared error (MSE) for $\eta$ estimation for the estimators introduced in this section (except $\widehat\eta_{\widetilde{M}}$) as a function of $\eta$, calculated from a Monte-Carlo simulation. These calculations were made for $c_{\mu} = 0.19~V$ and $c_{\sigma} = 0.0045 ~V^2$, which are close to the experimental values we obtained for these parameters as we will describe in \Cref{sec:experiment}. The oracle estimator sets a lower threshold for the MSE. At low $\eta$, the conventional and MLI estimators have similar MSEs, close to the ideal oracle performance. As discussed in \Cref{subsec:qm}, the QM estimator has a large bias at low $\eta$ and consequently a large MSE. For $\eta>1$, all estimators diverge significantly from the oracle, and the MLI estimator outperforms the conventional estimator. The QM estimator's MSE converges to that of the MLI estimator. This reduction in MSE would translate to lower noise in a PBM image, and it demonstrates the potential advantage in imaging performance possible with time-resolved measurement.

%\section{Conventional and ICA estimation of $\eta$ for bulk samples}
\subsection{Validation and comparison of estimators on bulk samples}
\label{subsec:etaexpt}
We used the measured $c_{\mu}$ to implement the conventional, QM, and ICA SE yield estimators. 
As discussed earlier, we used $\widehat\eta_{\Mtilde}$ as a benchmark for the other $\eta$ estimators at low $\eta$, where $\widehat\eta_{\Mtilde}$ is reliable. 
This benchmarking allowed us to be more confident in the $\eta$ estimators at high $\eta$, where $\widehat\eta_{\Mtilde}$ is not reliable. 
As discussed earlier, we collected 2.5\,s of data from featureless silicon samples, and we varied the effective $\eta$ by changing the SED collector bias. 
For the results described below, we used a pixel dwell time of $25\,\us$ and a beam current of $0.11\,\pA$, resulting in $\lambda = 17.2$ ions/pixel. 
For counting pulses and measuring their peak voltages, we used a threshold voltage of $0.07\,\V$ to filter out noise peaks.

%\AAcomment{Can we discuss notation here? I've used ICA, but perhaps that's too confusing. I do want the reader to remember that what we call CCMLI here is ICA in the paper.}

\begin{figure}
  \begin{center}
    \begin{tabular}{@{}c@{\,}c@{\,}}
      \includegraphics[width=0.49\linewidth]{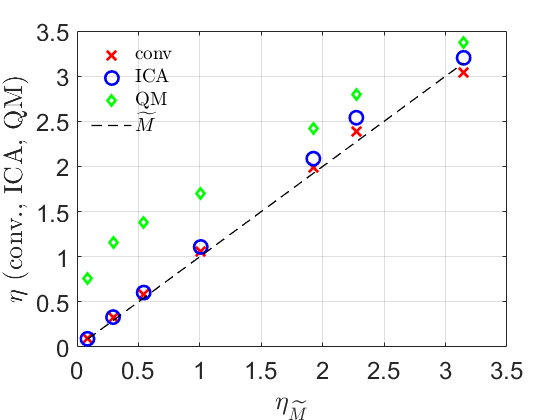} &
      \includegraphics[width=0.49\linewidth]{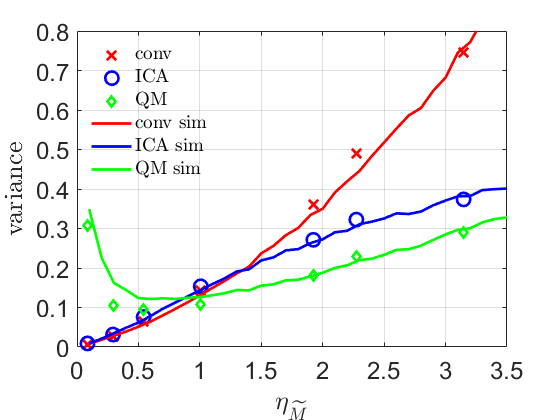} \\
      {\small (A) Mean vs.\ $\eta$} &
      {\small (B) Variance vs.\ $\eta$}
    \end{tabular}
  \end{center}
    \caption{Performance of ICA, QM, and conventional $\eta$ estimators for different values of $\eta$. The mean values of the ICA and the conventional estimators agree with the baseline $\widehat\eta_{\Mtilde}$, while the variance of the ICA estimator is lower than that of the conventional estimator for $\eta>1$. The QM estimator's variance is lower than the ICA, but it has a large bias.
    %\VKGcomment{Should the $\eta$ axis labels have hats?}
    %\AAcomment{I think so, because the $\Mtilde$ estimator is being used as the baseline.}
    %\VKGcomment{At the risk of being really annoying, should blue and red be reversed?}
    %\AAcomment{Fixed.}
    %\VKGcomment{Do you want to now change MLI to ICA in legends and axis labels?  (Check corresponding caption and body text, too, after this is decided.)}
    }
    \label{fig:estimator_performance}
\end{figure}

\Cref{fig:estimator_performance} shows the performance of the QM, ICA, and conventional estimators for $\eta$ values between 0 and 3.2\@. 
In \Cref{fig:estimator_performance}A, we compare the benchmark measurement of $\eta$, \textit{i.e.,} $\widehat\eta_{\Mtilde}$ (-\,-\,-), with
%We also indicate 95\% confidence intervals around $\widehat\eta_{\widetilde{M}}$, obtained from the statistics of $\widehat\eta_{\widetilde{M}}$ over all the pixels. 
%These intervals get wider at $\eta$ increases, indicating the increasing unreliability of $\widehat\eta_{\widetilde{M}}$.  
the conventional ({\color{blue} $\circ$}), ICA ({\color{red} $\times$}), and QM ({\color{green} $\diamond$}) estimates. 
We see that the conventional and ICA estimates are close to the benchmark values for the whole range of $\eta$.
It should be noted that $\widehat\eta_{\Mtilde}$ becomes increasingly unreliable as $\eta$ increases as discussed earlier, and therefore the differences between $\widehat\eta_{\Mtilde}$ and the other estimators at high $\eta$ might be due to inaccuracies in $\widehat\eta_{\Mtilde}$.
The QM estimator shows a significant bias at small $\eta$, as we would expect, and converges somewhat to the conventional and ICA estimators at higher $\eta$. 
The agreement between the benchmark and the conventional and ICA estimators is strong validation for our abstract model of SE generation.

\Cref{fig:estimator_performance}B shows the variance of the conventional, ICA, and QM estimators vs.\ $\eta$. 
The plot also shows theoretical curves for the variances of each of these estimators.
These curves were generated using Monte Carlo simulations of the signal produced by the SED using our model described at $\lambda = 17.2$ ions.
The voltage pulse from each burst of SEs was modelled as a Gaussian in the time domain, with FWHM of 160\,\ns.
Pulse heights and counts were extracted from this synthetic detector voltage provided as inputs to the $\eta$ estimators described in the previous sections.
The variances of the estimators are in line with theoretical expectations; the ICA estimator has lower variance than conventional for $\eta$ greater than $\sim1$. 
As an example, the variance for the ICA estimator is lower than that of the conventional by a factor of 2 at $\eta = 3$.
Since the variance is inversely proportional to imaging dose, we would expect a factor of 2 reduction in the imaging dose required for a given noise level compared to conventional imaging.
This factor of dose reduction is exactly what we had observed in the results presented in Figure 3 of the main paper.
The QM estimator has even lower variance than the ICA estimator at high $\eta$, but, as already discussed, has a relatively high bias, too. 
Overall, the lower variance of the ICA estimator is experimental evidence of its success in lowering noise in imaging.

\begin{figure}
  \begin{center}
    \begin{tabular}{@{}c@{\,}c@{\,}}
      \includegraphics[width=0.49\linewidth]{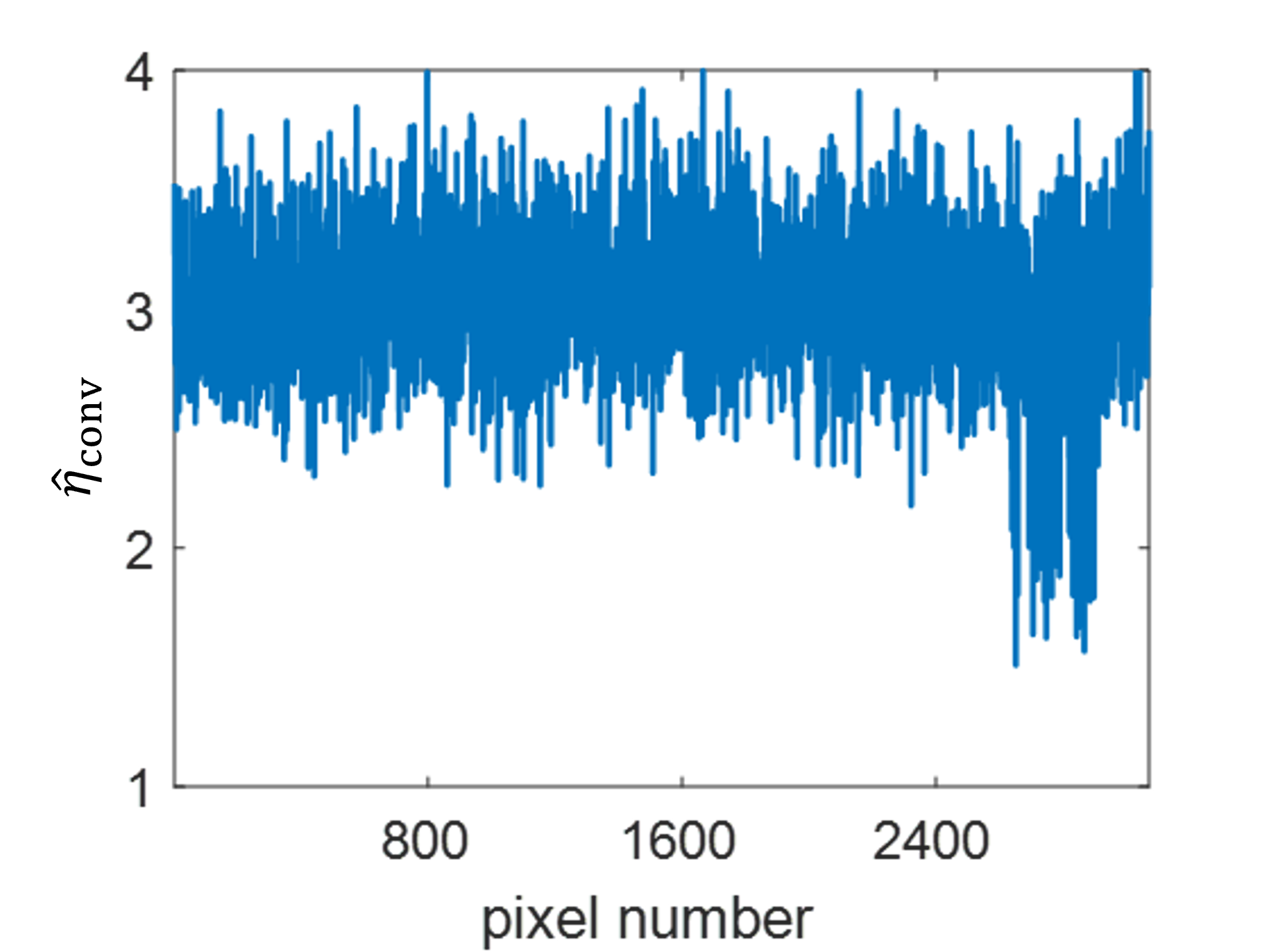} &
      \includegraphics[width=0.49\linewidth]{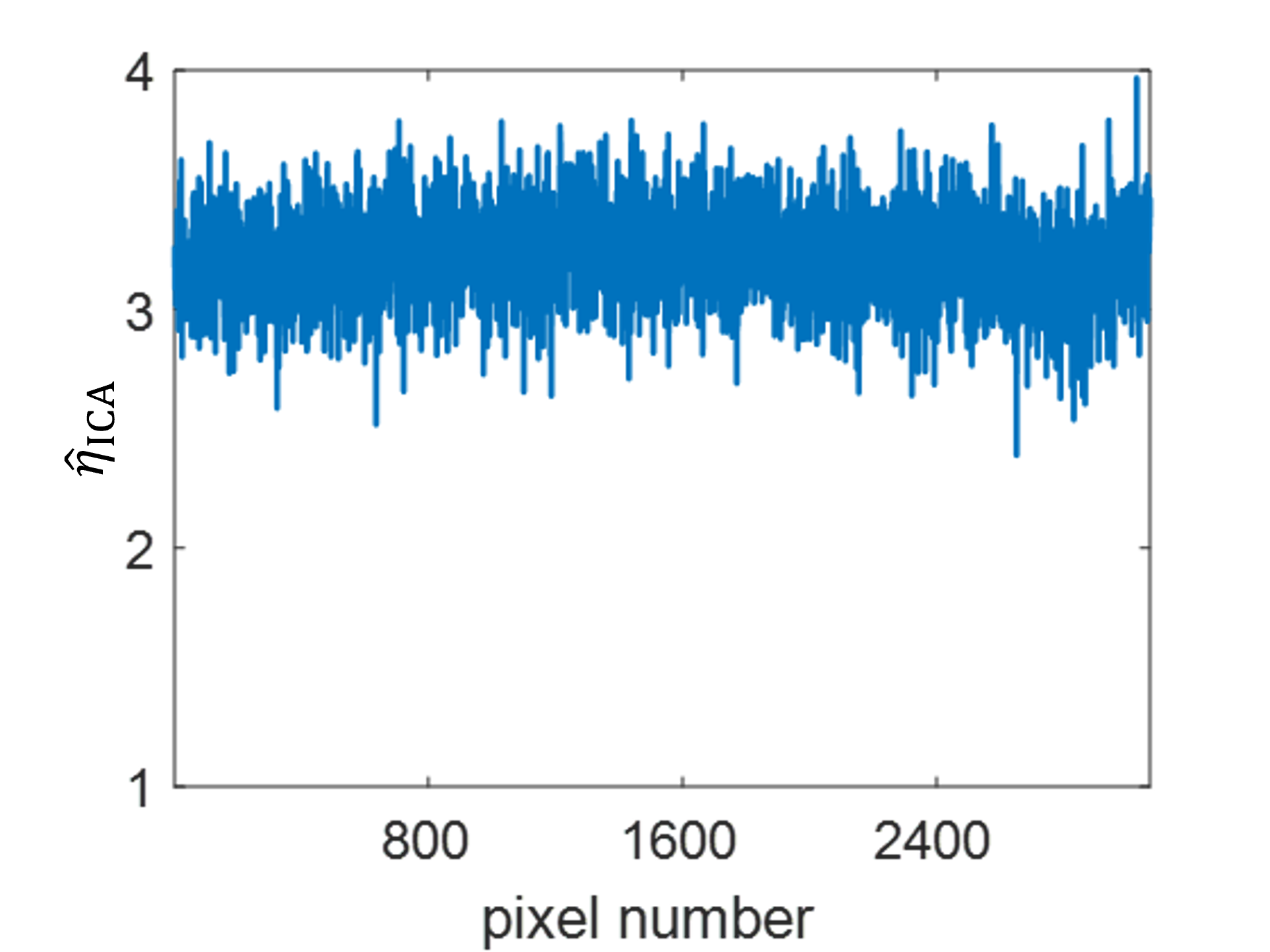} \\
      {\small (A) $\etaconv$} &
      {\small (B) $\etaICA$}
    \end{tabular}
  \end{center}
    \caption{One-dimensional imaging of a uniform silicon sample. The conventional estimate has higher variance (0.091) than the ICA estimate (0.036), resulting in a noisier reconstruction of $\eta$.
    %\VKGcomment{Fig S7B has ``trml''}
    }
    \label{fig:conv_trml_image}
\end{figure}

\Cref{fig:conv_trml_image} presents a different perspective on the reduction in variance from the ICA estimator.
%\AAcomment{This part and the figure accompanying it could be taken out. Doesn't add much.}
%\VKGcomment{I would like to keep it.  I can somehow imagine a certain sort of reader who is more convinced by this than by certain other computations.}
\Cref{fig:conv_trml_image}A is the conventional estimate of $\eta$ in a one-dimensional scan over the uniform silicon sample (at an SED voltage bias of 500 V). 
The mean value of $\eta$ here is 3.15\@. 
Since the sample is uniform, all variations in $\eta$ are due to randomness in $M$, $Y$, and $V$; the variance of the estimate is 0.091\@.  
\Cref{fig:conv_trml_image}B is the ion count-aided estimate of $\eta$ using the same dataset. 
%This estimate has almost the same mean value (3.15) as the conventional estimator. 
We can see that the noise in this estimate is lower than the conventional estimate; the variance of the ICA estimate is 0.036\@. 
Lower imaging noise results in increased resolution at the same dose.

\section{Similarity of Conventional Estimator and Software Images}
\label{subsec:conv_software}
%\VKGcomment{Moved this forward.  It is referenced quite early in the paper.}

\begin{figure}
    \centering
    \includegraphics[width = \linewidth]{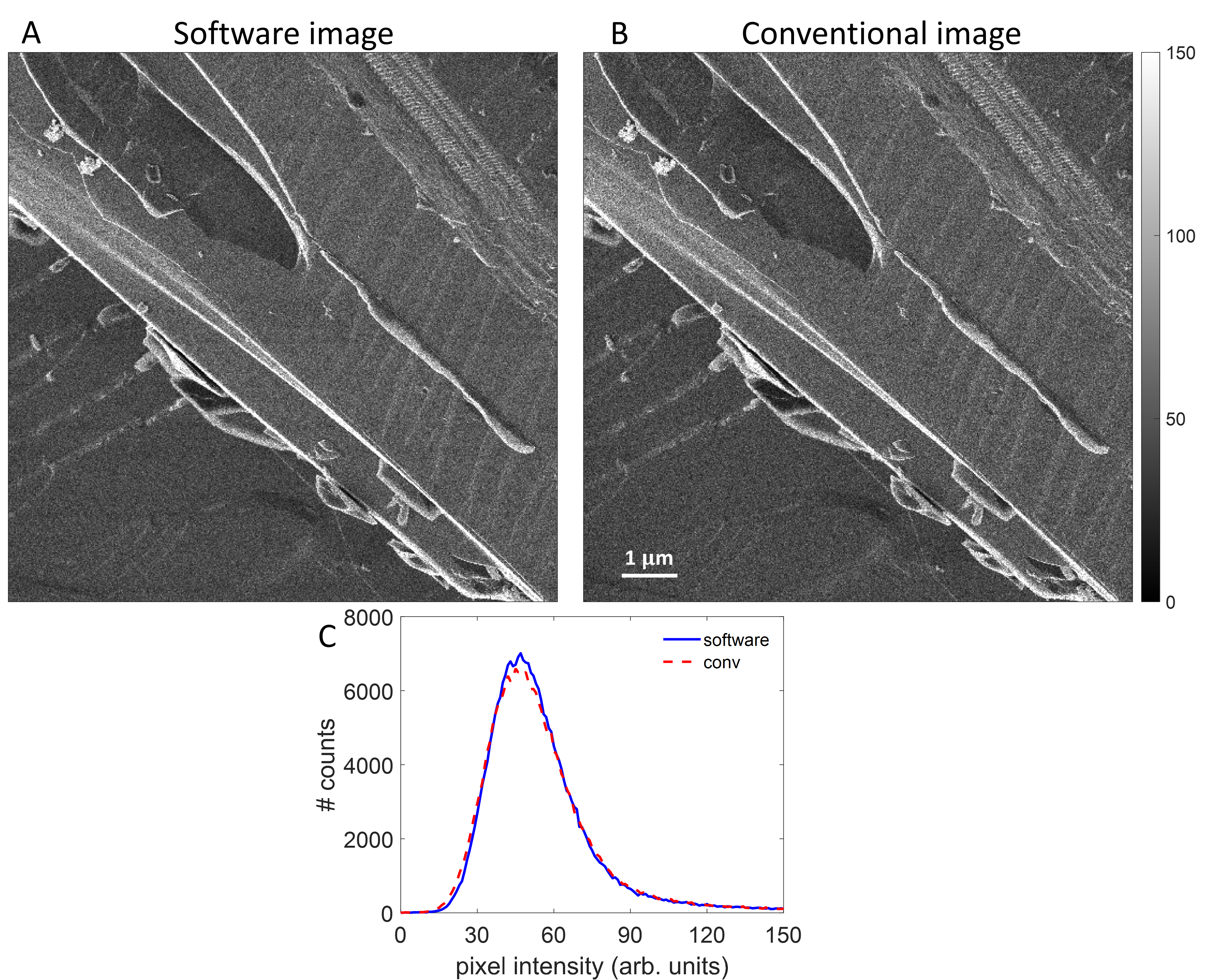}
    \caption{Conventional and software images of a scratch on silicon.
    (A) Image generated by microscope imaging software.
    (B) Image generated using conventional estimator. The images are visually similar.
    (C) Pixel intensity histograms of the two images. The two histograms are nearly identical.}
    \label{fig:conv_software_compare_1}
\end{figure}

\begin{figure}
    \centering
    \includegraphics[width = \linewidth]{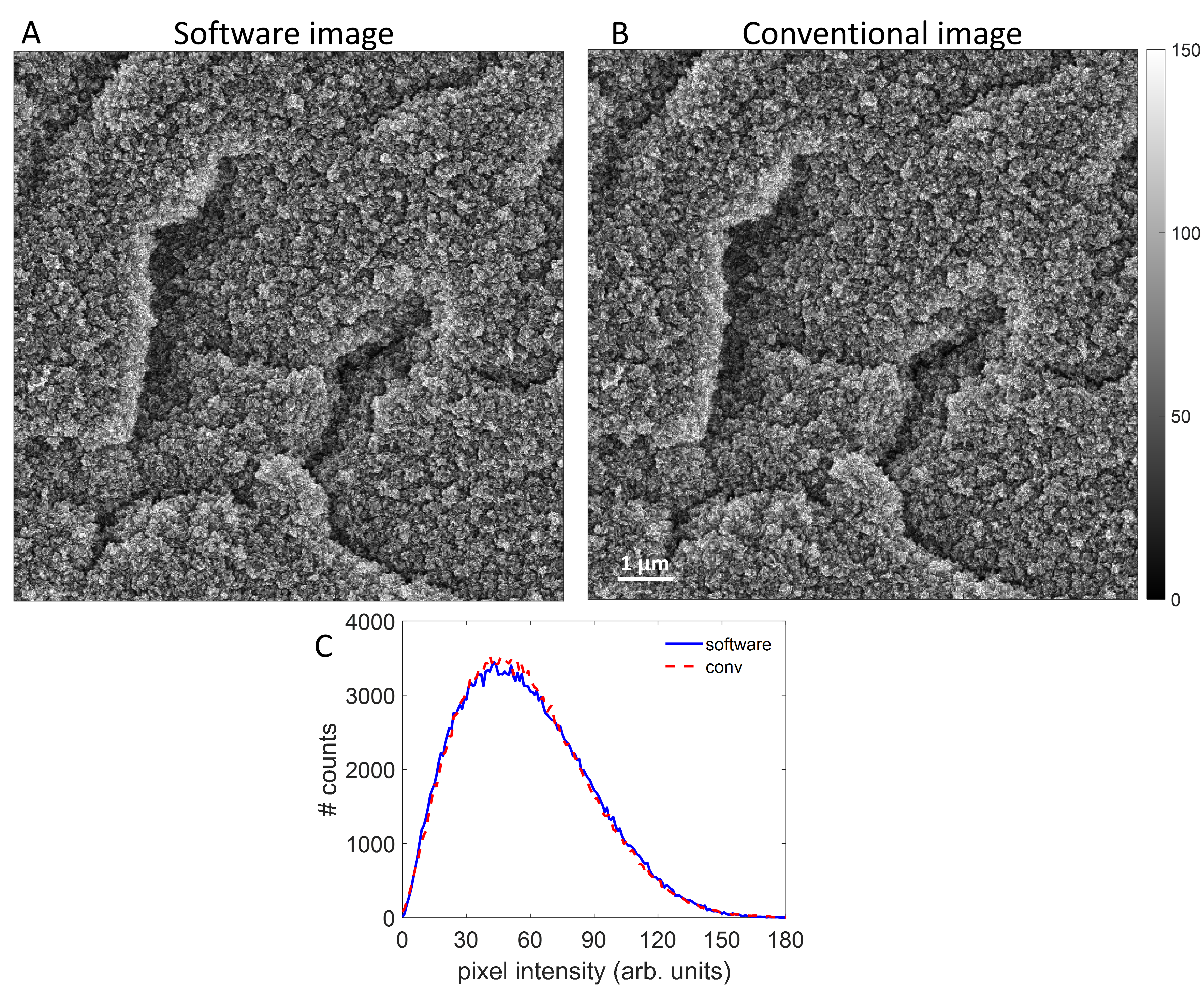}
    \caption{Conventional and software images of agglomerated silver nanoparticles.
    (A) Image generated by microscope imaging software.
    (B) Image generated using conventional estimator. The images are visually similar.
    (C) Pixel intensity histograms of the two images. The two histograms are nearly identical.}
    \label{fig:conv_software_compare_2}
\end{figure}

As discussed in the main paper, we used the conventional estimator to represent typical SEI\@. 
\Cref{fig:conv_software_compare_1,fig:conv_software_compare_2} show two examples of images produced by our conventional estimator alongside images produced by the SEI software on the microscope. 
For the purposes of this comparison, we scaled the conventional image to have the same mean as the software image, since the two images are originally on different scales.
\Cref{fig:conv_software_compare_1} compares the (A) software and (B) conventional images of the same sample that was used in Figure 2 in the paper.
Note that due to the rescaling of the conventional image, the grayscale does not reflect SE yields anymore.
We can see that the two images appear to be very similar.
The visual similarity is confirmed by the image histograms of the two images in \Cref{fig:conv_software_compare_1}C.
\Cref{fig:conv_software_compare_2} compares the software and conventional images of a sample of agglomerated silver nanoparticles.
Again, the two images and their histograms are nearly identical.

\section{Uncropped Images for Figure 2}
\label{subsec:Full_image}
\begin{figure}
    \centering
    \includegraphics[width = \linewidth]{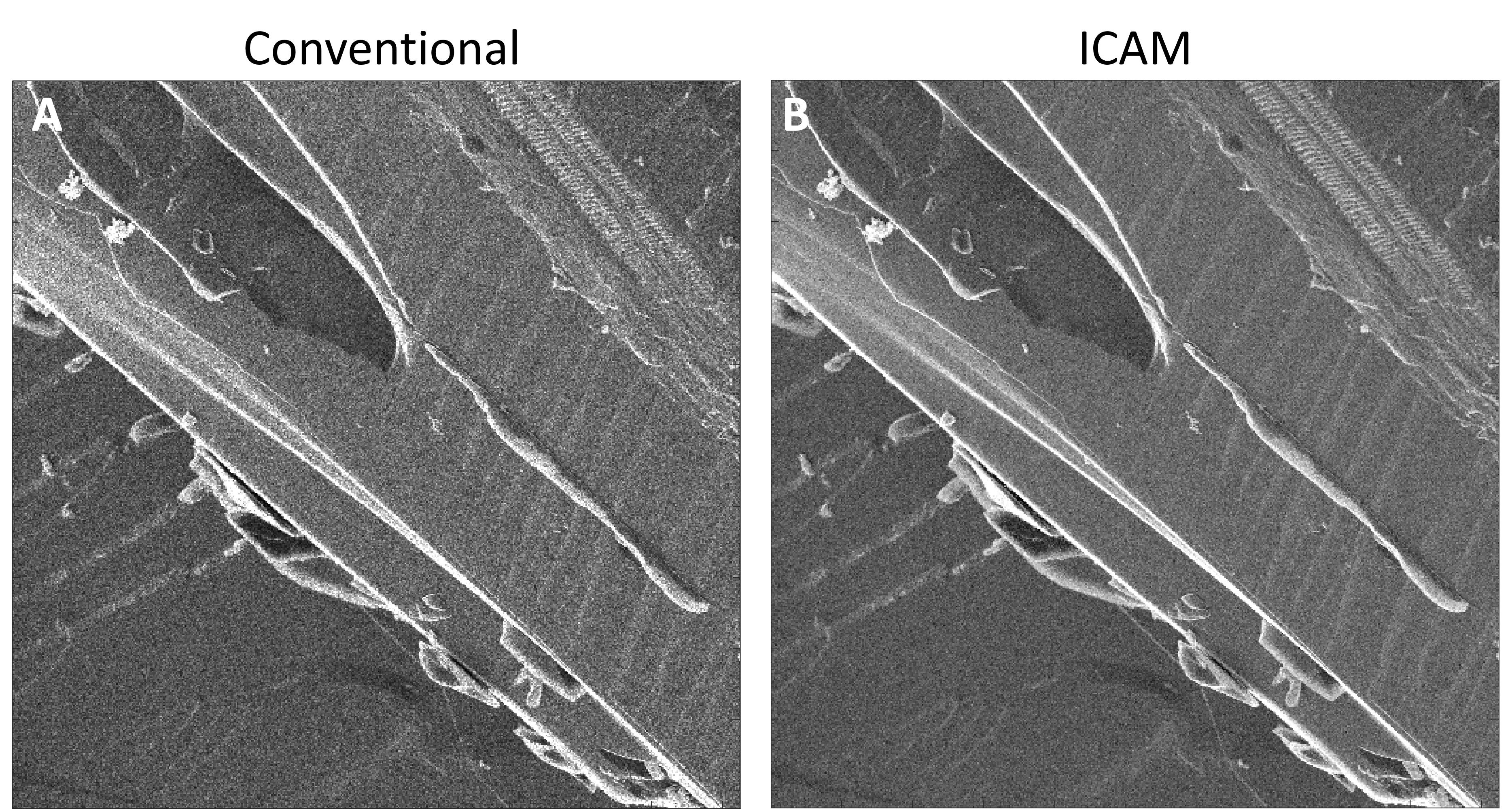}
    \caption{Uncropped images of the sample imaged in Figure 2 at dose $\lambda = 22$ ions. (A) Conventional image (B) ICAM image.}
    \label{fig:uncropped}
\end{figure}

\Cref{fig:uncropped} shows uncropped conventional and ICAM images of the same silicon sample used to create Figure 2 in the paper.

\section{Performance of conventional and ICAM imaging at high magnification}
\label{subsec:high_mag}
\begin{figure}
    \centering
    \includegraphics[width = \linewidth]{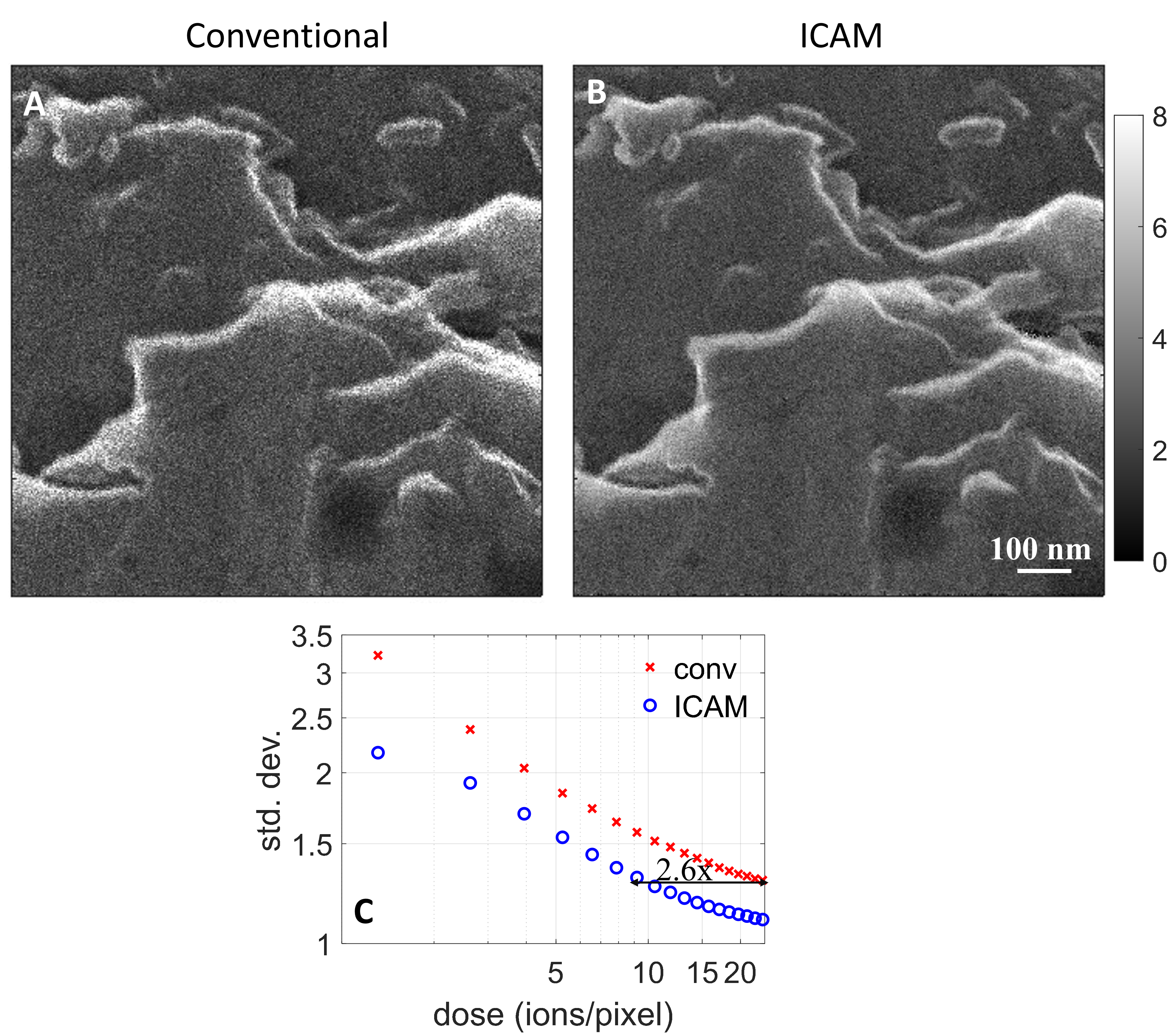}
    \caption{$1~\um$ field-of-view image of silicon sample at dose $\lambda = 22$ ions. (A) Conventional image. (B) ICAM image. (C) Comparison of image standard deviation vs dose for conventional and ICAM imaging.}
    \label{fig:high_mag}
\end{figure}

\Cref{fig:high_mag}A and B shows an exmaple of conventional and ICAM imaging at a field of view of $1~\um$, respectively. \Cref{fig:high_mag}C is a plot of the image standard deviation as a function of dose for the two imaging modes. We can see that ICAM achieves reduction in dose (between a factor of 2 and 3) for a given standard deviation. This gain is similar to that reported in the paper at a field of view of $10~\um$. This result, along with the agreement between theoretical and experimental predictions in Figure 4 of the paper, show that the gains offered by ICAM are independent of image magnification.

\section{Performance Metrics for Conventional and ICAM Images}
\label{subsec:SNR_FRC}
\begin{figure}[ht]
    \centering
    \includegraphics[width = \linewidth]{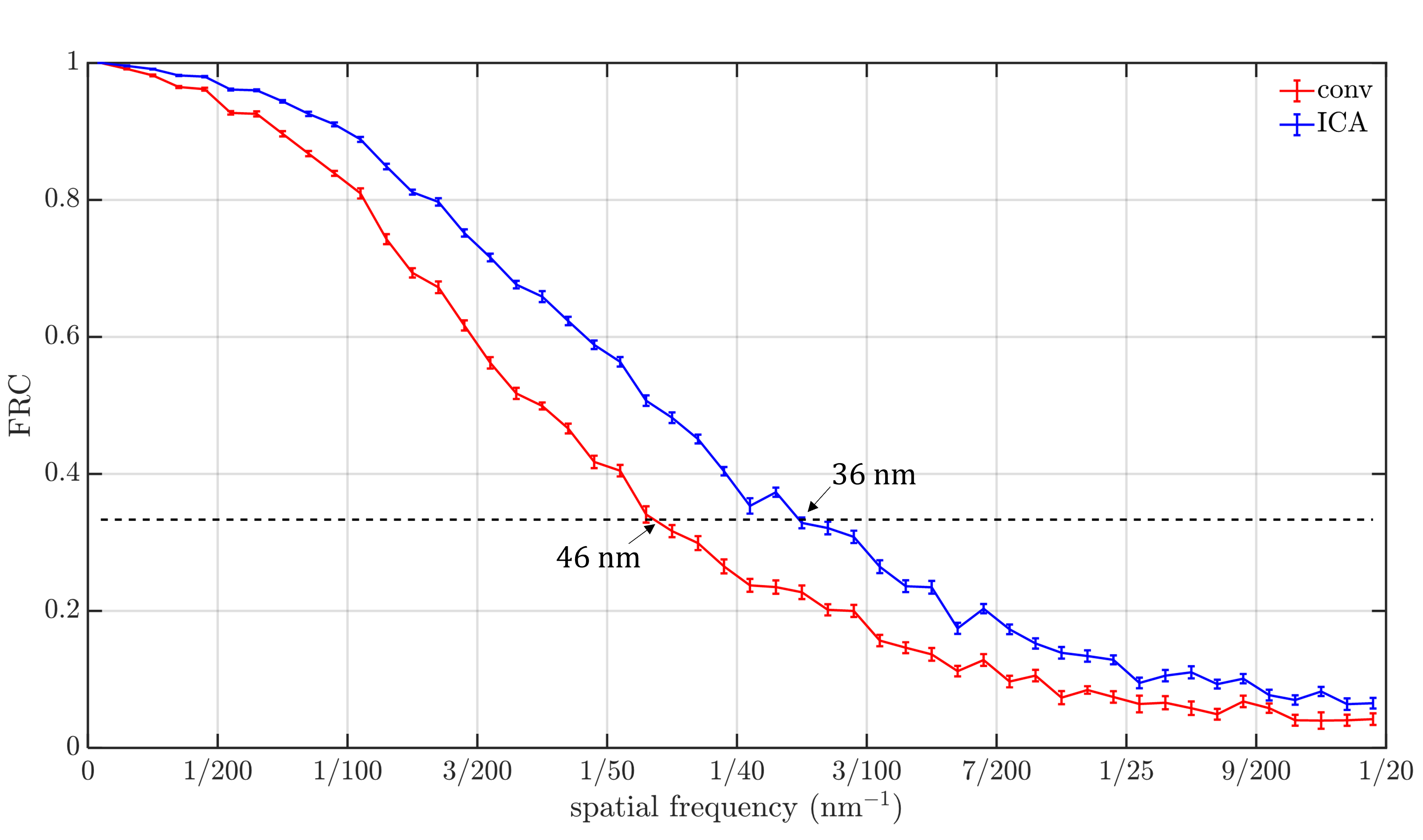}
    \caption{Fourier Ring Correlation for conventional (red) and ICA (blue) estimators. The 1/3 criterion resolution is $46\,\si{\nm}$ for conventional and $36\,\si{\nm}$ for ICAM.}
    \label{fig:frc}
\end{figure}

\textbf{Silicon scratch sample}: In the paper we calculated Thong's SNR metric for the conventional and ICAM images of the silicon scratch sample.
In addition to this metric, we can also calculate SNR using Saxton's method~\cite{Saxton1978}. 
Saxton proved that the SNR could be computed from the normalized cross-correlation of two independently acquired, zero-mean images of a ground truth sample.
To implement this SNR metric, we divided our dataset into two halves, and we created conventional and ICAM images from each half-dataset independently.
Next, we computed the normalized cross-correlation at zero offset between these half-dataset images after subtracting their means, and used the result to compute the SNR\@.
We repeated this calculation for 1000 different half-dataset divisions of the full dataset.
We obtained an average SNR of 1.17 for the conventional image, and 2.11 for the ICAM image, again indicating an SNR improvement by a factor of 2\@.

The Fourier Ring Correlation (FRC)~\cite{vanHeel1982,Saxton1982,vanHeel1987a} is a resolution metric commonly used in cryo-electron microscopy~\cite{Rosenthal2003}. 
%Recently, FRC has also been used in optical microscopy~\cite{Banterle2013,Lidke2013}. 
This metric attempts to find the highest spatial frequency at which the Fourier spectrum of the image contains useful signal by looking at the normalized cross-correlation between the Fourier transforms of two half-dataset images, computed within rings of increasing spatial frequency.
The resolution is defined as the inverse of the spatial frequency at which the FRC is greater than a pre-decided threshold.
A number of different thresholds are used for defining resolution~\cite{Rosenthal2003,Sorzano2017}.
Here, we will use 1/3 threshold.
Similar to our calculation of Saxton's SNR metric, we repeated our calculation of the FRC for 1000 independent half-dataset divisions.
\Cref{fig:frc} shows the average FRC curves for the conventional (red curve) and ICA (blue curve) estimators, along with error bars that correspond to the standard deviation of each FRC value over the 1000 dataset divisions.
The 1/3 resolution criterion is indicated by the horizontal black dashed line.
The resolution is {46\,\si{\nm}} for the conventional estimator and {36\,\si{\nm}} for the ICA estimator.

\textbf{Gold on silicon sample}:
In the paper, we characterized this sample by calculating Thong's SNR, as well as the dose reduction for the silicon and gold regions. 
We can further exploit the binary nature of this sample to define a distinguishability metric $D$ between the bright and dark levels as
%\AAcomment{This was a metric suggested by John Notte that I thought was an interesting way to quantify the binary sample.}
\begin{equation*}
        D = \frac{|\eta_1-\eta_2|}{\sqrt{\sigma_1 \sigma_2}},
\end{equation*}
where $\eta_1$ and $\eta_2$ are the mean SE yields and $\sigma_1$ and $\sigma_2$ are the standard deviations of the SE yield values for the two pixel types.
A higher value of this metric would imply greater ability to distingush between gold and silicon pixels, and consequently, higher image quality.
We get $D=1.87$ for the ICAM image and $D=1.50$ for the conventional image.

\section{Calculation of SED Detector Quantum Efficiency}
\label{subsec:DQE}
As discussed briefly in the paper, our measured SE yield values are not absolute, since we did not separate the effect of non-ideal detective quantum efficiency (DQE) from the calculation. 
The DQE is defined as the fraction of emitted SEs that produce a signal on the SED~\cite{Joy2008, AGARWAL2021}. 
Since the DQE of the SED can depend on several factors, such as the relative placement of the sample and the SED, the working distance, and the type of detector (in-lens vs.\ in-chamber), we only quoted the as-measured SE yield values in the paper.

\begin{figure}
    \centering
    \includegraphics[width = \linewidth]{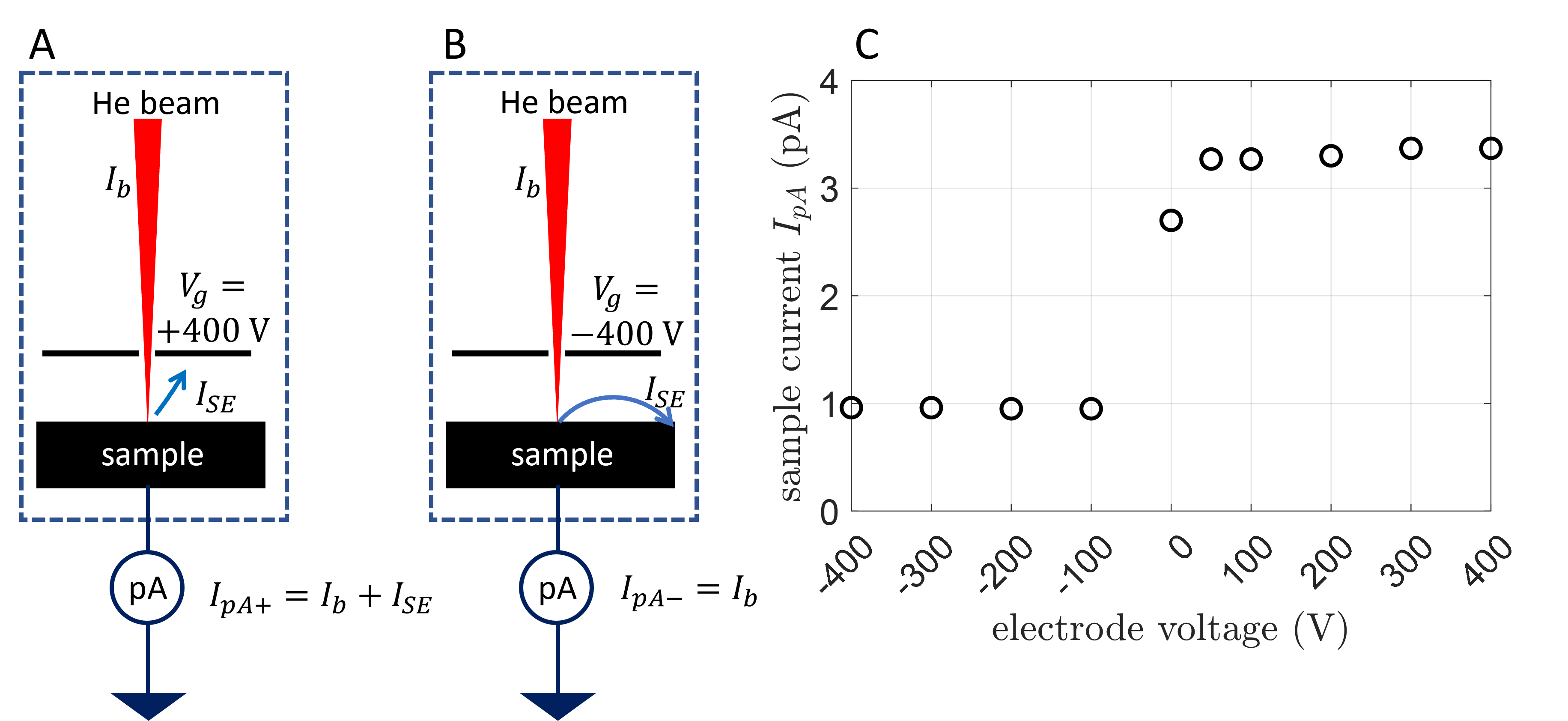}
    \caption{Measurement of bulk sample SE yield.
    (A) When electrode voltage $V_g$ is highly positive, emitted SEs are drawn away from the sample. The total measured sample current is the sum of the beam and SE currents.
    (B) When $V_g$ is highly negative, emitted SEs are repelled back towards the sample. The measured sample current is just the beam current.
    (C) Plot of sample current as a function of electrode voltage.
    %\VKGcomment{There are some inconsistencies between figure and body text.  Do you want to put $I_{pA+}$ and $I_{pA-}$ in the figure?  In C, the yaxislabel should have $I_{pA}$ instead of pA.}
    %\AAcomment{Fixed some notational issues. `pA' is pulling double duty here. It both stands for the instrument being used to measure the current, the picoammeter, and the unit of the current, picoAmperes. }
    %\VKG{Oh!  I didn't notice that.}
    }
    \label{fig:DQE_setup}
\end{figure}

Nevertheless, the DQE for a particular experimental geometry can be measured. 
Our measurement of the SE yield through the methods described in the paper are equivalent to measuring the average number of detected SEs. 
If we can independently measure the average number of emitted SEs, we can calculate the DQE\@.
For this purpose, we used the setups depicted in \Cref{fig:DQE_setup}, A and B\@. 
We placed an annular copper electrode above a featureless silicon sample, and we measured the sample current $I_{pA}$ at various electrode voltages using a picoammeter.
When the electrode voltage $V_g$ is highly positive, as depicted in \Cref{fig:DQE_setup}A, we expect $I_{pA}$ to be the sum of the beam current $I_b$ and the SE current $I_{SE}$. 
When $V_g$ is highly negative, SEs are repelled away from the electrode and back to the sample, and $I_{pA}$ will be equal to $I_b$.
Therefore, the SE yield is given by $\widehat{\eta} = (\Frac{I_{pA+}}{I_{pA-}})-1$.
\Cref{fig:DQE_setup}C is a plot of $I_{pA}$ as a function of the applied electrode voltage, measured at $I_b = 1\,\pA$. 
We can see that, as expected, when $V_g$ is highly negative, $I_{pA} = I_b$, and when the grid voltage is highly positive, $I_{pA}$ is higher than $I_b$. 
From these measurements, we extract $\widehat{\eta} = 2.51$.
For this sample, the ICA estimator gave $\widehat{\eta} = 2.20$.
From the ratio between these numbers, we get DQE = 88\%.

\section{Beam Current Sensitivity of Conventional and ICA Estimators}
\label{subsec:current_sensitivity}
From the expressions for the estimators, we can compute their sensitivities to changes in the incident current (equivalently, changes in dose $\lambda$).
From \cref{eqn:conv}, it is straightforward to conclude that
\begin{equation*}
    \frac{\partial \etaconv}{\etaconv} = - \frac{\partial \lambda}{\lambda}.
\end{equation*}
Therefore, 10\% fluctuation in $\lambda$ causes the the conventional SE yield estimate to fluctuate by 10\%. 
For the ICA estimator, we can use \cref{eqn:mli} to write
\begin{align*}
    \frac{\partial \etaICA}{\etaICA} & = - \frac{\partial \lambda}{\lambda}\left(\frac{e^{-\etaICA}}{\Mtilde/\lambda - e^{-\etaICA}(\etaICA-1)}\right) \\
    & \approx -\frac{\partial \lambda}{\lambda}\left(\frac{1}{e^{\etaICA} - \etaICA}\right).
\end{align*}
%\AAcomment{Should notation be MLI or ICA?}
%\VKGcomment{I like the choice of MLI because ICA is really CCMLI, which has the complication of $\gamma$.}
Here the first equality follows from differentiating \cref{eqn:mli} with respect to $\lambda$.
We approximate $\Mtilde$ by its mean value of $\lambda(1-e^{-\eta})$ to get the final expression. 
At $\eta = 2.75$, a 10\% fluctuation in $\lambda$ causes the ICA SE yield estimate to fluctuate by only $\sim0.8\%$.

\section{Relative Error Dependence on Scaling of SE Yield}
\label{subsec:heavier_ions}
The use of ICAM in SEI creates quantitative estimates of SE yield $\eta$, replacing the qualitative nature of most SEI\@.
Nevertheless, there are possibilities for scaling of $\eta$ that prompt us to comment on how the size of $\eta$ affects imaging performance.
For example, varying SED collector bias will change the effective $\eta$ (see \Cref{subsec:etaexpt}),
and replacing helium with heavier ions at the same velocity will generally increase $\eta$~\cite{Fehn1976}.
We argue below that ICAM makes having large values of $\eta$ favorable,
whereas conventional imaging gives little to no benefit for large $\eta$.

\begin{figure}
    \centering
    \includegraphics[width=0.6\linewidth]{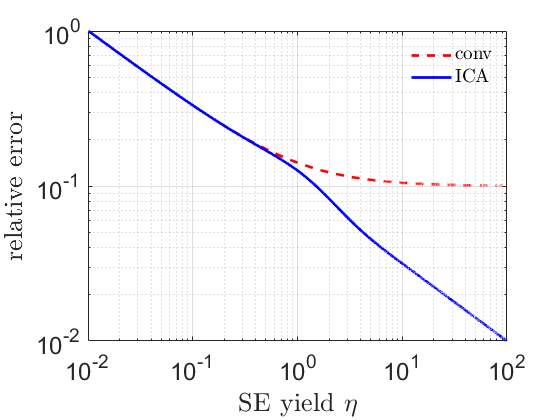}
    \caption{Comparison of relative errors for conventional and ICA estimation [see \cref{eq:REconv,eq:REICA}].}
    \label{fig:RE-comparison}
\end{figure}

For simplicity, let us assume noiseless SED ($c_\sigma = 0$).  Then \cref{eq:conventional_MSE} gives
\begin{equation}
    \MSEconv = \frac{\eta(\eta+1)}{\lambda}.
\end{equation}
As a scale-invariant accuracy metric,
consider \emph{relative error} defined by normalizing the root mean-squared error by the quantity of interest:
%\AAcomment{Should ``relative error'' be italicized or emphasized on in some way? At first read it isn't clear that the above sentence is defining relative error.}
\begin{equation}
\label{eq:REconv}
    \REconv = \frac{ \sqrt{\MSEconv} }{ \eta }
            = \frac{ \sqrt{\eta(\eta+1)} }{ \eta \sqrt{\lambda} }
            = \frac{ \sqrt{1 + 1/\eta} }{ \sqrt{\lambda} }.
\end{equation}
(Since the estimator is unbiased, this quantity may be familiar as the \emph{coefficient of variation}.)
For comparison, in the idealized case of noiseless SED, we can expect the performance of ICAM to follow predictions from Fisher information calculations~\cite{PengMBG:21}.
These suggest
\begin{equation}
    \MSEICA = \frac{\eta}{\lambda(1-\eta e^{-\eta})},
\end{equation}
which yields relative error
\begin{equation}
\label{eq:REICA}
    \REICA = \frac{ \sqrt{\MSEICA} }{ \eta }
           = \frac{ (1 - \eta e^{-\eta})^{-1/2} }{ \sqrt{\eta} \sqrt{\lambda} }.
\end{equation}
%\VKGcomment{Is there a more easily interpretative way to write that?}
To compare \cref{eq:REconv,eq:REICA},
first note that the numerators satisfy
\begin{equation}
           { (1 - \eta e^{-\eta})^{-1/2} }
           < 
           { \sqrt{1 + 1/\eta} }
           \qquad
           \mbox{for all $\eta > 0$},
\end{equation}
with
\begin{equation}
    \lim_{\eta \rightarrow \infty} 
           { (1 - \eta e^{-\eta})^{-1/2} }
           = 1
           \qquad \mbox{and} \qquad
    \lim_{\eta \rightarrow \infty} 
           { \sqrt{1 + 1/\eta} }
           = 1.
\end{equation}
Additionally, the denominator of \cref{eq:REICA} shows that $\REICA$ decays as $\eta^{-1/2}$ with increasing $\eta$.
The relative errors are plotted in \Cref{fig:RE-comparison} for $\lambda = 100$.
Notice that the relative error for conventional image formation approaches a floor of $\lambda^{-1/2} = 0.1$
whereas the relative error for ICAM is always lower and exhibits the expected $\eta^{-1/2}$ decay
(slope of $-1/2$ on a log-log plot).

This analysis shows that the computational methods for image formation generate significantly different behaviors:
under conventional image formation, there is little to no benefit to increasing $\eta$, whereas ICAM produces images of better accuracy when $\eta$ is increased.
%In the context of choosing among incident particle types \VKG{with the velocity kept fixed},
%with conventional image formation,
%heavier particles produce more sample damage,
%but \AAtext{\sout{apparently}} without
%%the increased SE yield $\eta$ is not
%improving image quality;
%ICAM would apparently improve the image quality,
%creating a new trade-off between image quality and sample damage.
The improvement in relative error with increasing $\eta$ creates a new trade-off between image quality and sample damage that could change the incident particle type and energy for optimal imaging resolution.
%\AAcomment{Maybe we can say something like ``The improvement in relative error with increasing $\eta$ creates a new trade-off between image quality and sample damage which could change the incident particle type and energy for optimal imaging resolution.'' In this formulation we're not saying that heavier particles necessarily lead to higher SE yield.}
% Start a new page so that we can discard the references from the PDF file.
\clearpage

~
%\newpage

%\bibliographystyle{ScienceAdvances}
\bibliographystyle{ieeetr}
\bibliography{bibliograph}